\documentclass[journal]{IEEEtran}
\usepackage{graphics} 
\usepackage{epsfig} 
\usepackage{times} 
\usepackage{amsmath} 
\usepackage{amssymb}  
\usepackage{cite}
\usepackage{epstopdf}
\usepackage{subfigure,color,balance}
\usepackage{verbatim}
\usepackage{cases}
\usepackage{enumerate}
\usepackage{booktabs}
\usepackage{balance}
\usepackage{threeparttable}
\usepackage{mathbbol}
\usepackage{algorithm}
\usepackage{algorithmicx}
\usepackage{algpseudocode}

\newcommand\markend{{\hfill {\small$\square$}}}

\usepackage[colorlinks=true,      
linkcolor=black,      
citecolor=black,      
filecolor=black,      
urlcolor=blue]{hyperref}

\newtheorem{definition}{Definition}
\newtheorem{theorem}{Theorem}

\newtheorem{remark}{Remark}
\newtheorem{lemma}{Lemma}

\newtheorem{problem}{Problem}

\newtheorem{conjecture}{Conjecture}
\newtheorem{fact}{Fact}
\newtheorem{example}{Example}

\def\R{{\bf R}}

\def\0{{\bf 0}}
\def\1{{\bf 1}}

\def\eg{{\em e.g.}}
\def\ie{{\em i.e.}}

\newcommand{\tr}{{{\mathsf T}}}

\begin{document}
%
\title{
Cooperative Formation of Autonomous Vehicles in Mixed Traffic Flow: Beyond Platooning}

\author{Keqiang Li, Jiawei Wang,~\IEEEmembership{Student Member,~IEEE,} and Yang Zheng,~\IEEEmembership{Member,~IEEE}

\thanks{The work of K.~Li and J.~Wang is supported by National Key R\&D Program of China with 2018YFE0204302 and Key-Area R\&D Program of Guangdong Province with 2019B090912001. K. Li and J. Wang contributed equally to this work. (Corresponding author: Y. Zheng)}
\thanks{K.~Li and J.~Wang are with the School of Vehicle and Mobility, Tsinghua University, Beijing, China, and with the Center for Intelligent Connected Vehicles \& Transportation, Tsinghua University, Beijing, China (email: likq@tsinghua.edu.cn,wang-jw18@mails.tsinghua.edu.cn).}%
\thanks{Y. Zheng is with the Department of Electrical and Computer Engineering, University of California San Diego, CA, USA. ({zhengy@eng.ucsd.edu}).}%
}

\maketitle

\begin{abstract}
	Cooperative formation and control of autonomous vehicles (AVs) promise increased efficiency and safety on public roads. In single-lane mixed traffic consisting of AVs and human-driven vehicles (HDVs), the prevailing platooning of multiple AVs 
is not the only choice for cooperative formation. In this paper, we investigate how different formations of AVs impact traffic performance from a set-function optimization perspective. We first reveal a stability invariance property and a diminishing improvement property of noncooperative formation when AVs adopt 
an independently designed Adaptive Cruise Control (ACC) strategy. Then, we focus on the case of cooperative formation where AVs utilize a centralized optimal controller. We further investigate the corresponding optimal 
formation of multiple AVs using set-function optimization. Two predominant optimal formations, \textit{i.e.}, uniform distribution and platoon formation, emerge from extensive numerical experiments. Interestingly, platooning might have the least potential to improve traffic performance when HDVs have poor string stability behavior. 
These results suggest more opportunities for cooperative formation of AVs, beyond platooning, in practical mixed traffic flow.

\end{abstract}


%
\IEEEpeerreviewmaketitle

\section{Introduction}

\IEEEPARstart{R}{educing} traffic congestion and achieving better mobility have received significant interest 
since the popularization of automobiles in the early 20th century. For a series of human-driven vehicles (HDVs), it is known that small perturbations may lead to stop-and-go waves, propagating upstream traffic flow~\cite{sugiyama2008traffic}. This phenomenon of traffic instability, known as \emph{phantom traffic jam}, can result in a great loss of travel efficiency and fuel economy~\cite{batty2008the}. The emergence of autonomous vehicles (AVs) is expected to smooth traffic flow and improve traffic efficiency, as the motion of AVs can be directly controlled. 
In particular, cooperative formation and control of multiple AVs promise to revolutionize road transportation systems in the near future~\cite{xu2021coordinated}.

\subsection{Formation and Control of Multiple AVs}

Platooning is one typical formation of multiple AVs, attracting significant attention in the past decades~\cite{li2017dynamical,zheng2015stability,ploeg2014controller,amoozadeh2015platoon,menaoreja2018permit,mena2018impact,deng2016general}. In a platoon formation, adjacent vehicles are regulated to maintain the same desired velocity while keeping a pre-specified inter-vehicle distance. The earliest practice of platooning dates back to the PATH program in the 1980s~\cite{shladover1991automated}, followed by other famous programs around the world, including GCDC in the Netherlands~\cite{kianfar2012design}, SARTRE in Europe~\cite{robinson2010operating}, and Energy-ITS in Japan~\cite{tsugawa2011automated}. Both theoretical analysis~\cite{li2017dynamical,zheng2015stability,ploeg2014controller} and real-world experiments~\cite{shladover1991automated,kianfar2012design,robinson2010operating,tsugawa2011automated} have confirmed the great potential of vehicle platooning in achieving higher traffic efficiency, better driving safety, and lower fuel consumption. As the gradual deployment of AVs, however, there will be a long transition phase of mixed traffic flow, where both AVs and HDVs coexist. This brings a challenge for practical implementation of vehicle platooning. Since AVs are usually distributed randomly in real traffic flow---a sparse and random distribution is common at a low penetration rate, several maneuvers including joining, leaving, merging, and splitting need to be performed to organize neighboring AVs into a platoon; see, \eg,~\cite{amoozadeh2015platoon,menaoreja2018permit}. These maneuvers might bring possible negative impacts, even causing undesired congestion~\cite{mena2018impact,deng2016general}. These results suggest reconsidering the necessity of forming a platoon of multiple AVs in the mixed traffic environment.

\begin{figure}[t]
	\centering
	\subfigure[Uniform distribution]
	{
		\label{Fig:Formationillustration_Unifrom}
		\includegraphics[scale=0.18]{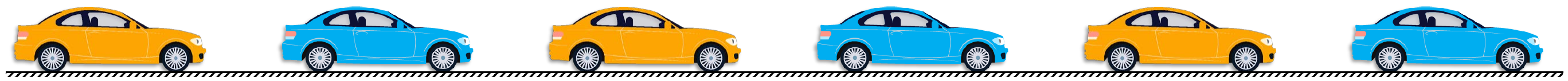}
	}
	\subfigure[Random formation]
	{ \label{Fig:Formationillustration_Random}
		\includegraphics[scale=0.18]{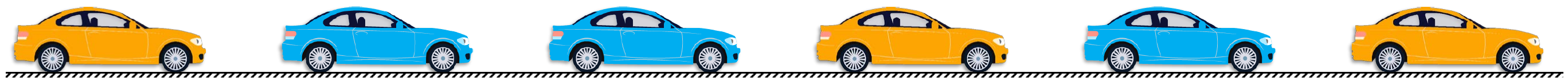}
	}
	\subfigure[Platoon formation]
	{\label{Fig:Formationillustration_Platoon}
		\includegraphics[scale=0.18]{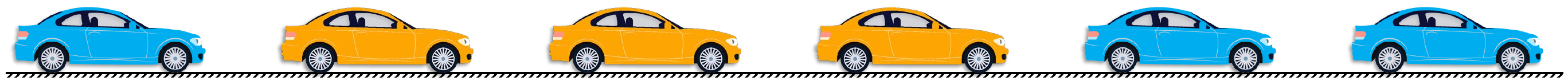}
	}
	\vspace{-2mm}
	\caption{Three examples for possible formations of AVs in mixed traffic flow. Blue vehicles and yellow vehicles represent HDVs and AVs, respectively.}
	\label{Fig:FormationIllustration}
	\vspace{-5mm}
\end{figure}

In fact, platooning is not the only formation of AVs in mixed traffic flow. Possible choices can be more diverse since AVs need not to drive in a consecutive manner in mixed traffic. 
For example, uniform distribution (see Fig.~\ref{Fig:Formationillustration_Unifrom}) or random formation (see Fig.~\ref{Fig:Formationillustration_Random}) of AVs could be possible options, besides the prevailing platoon formation (see Fig.~\ref{Fig:Formationillustration_Platoon}). A closely relevant concept is \emph{spatial distribution}, and the influence of the spatial distribution of AVs has been recently investigated via theoretical analysis~\cite{xie2019heterogeneous} and traffic simulation~\cite{jin2020impact}. However, most existing research considered noncooperative controllers for AVs, \eg, adopting a typical Adaptive Cruise Control (ACC) strategy~\cite{xiao2010comprehensive} that is locally designed with no cooperation. The potential of centralized cooperative control for AVs has been neglected in~\cite{xie2019heterogeneous,jin2020impact}. This class of formations with individually designed controller is called as \emph{noncooperative formation}, and we refer to \emph{cooperative formation} as a spatial distribution maintained by AVs using centralized cooperative control in mixed traffic flow. Given a specific formation of AVs in mixed traffic, \eg, platoon formation or random formation, the topic of designing cooperative control strategies for AVs has also received increasing interest, and a variety of methods have been introduced, including model-based strategies~\cite{di2019cooperative,gong2018cooperative} and data-driven strategies~\cite{vinitsky2018lagrangian,wu2017flow} (see~\cite{li2017dynamical} and~\cite[Section 6.1]{di2020survey} for recent surveys). 
It remains unclear which formation of AVs could achieve a better system-wide performance for mixed traffic flow.



Our main focus is to investigate the role of cooperative formation in improving traffic performance, and identify the optimal formation pattern for AVs in mixed traffic. 
Specifically, we utilize the notion of \emph{Lagrangian control} of traffic flow~\cite{stern2018dissipation} to achieve centralized cooperative control for AVs. One key idea is to employ AVs as \emph{mobile actuators} for traffic control through their direct interaction with neighboring vehicles. The effectiveness of this notion in reducing traffic instabilities and smoothing traffic flow has been investigated in the case of one single AV; see recent rigorous theoretical analysis~\cite{zheng2020smoothing,cui2017stabilizing,wang2020controllability}, small-scale real-world experiments~\cite{stern2018dissipation} and large-scale numerical simulations~\cite{vinitsky2018lagrangian}. Along this direction, it is natural to consider the case with multiple AVs coexisting, where the mixed traffic flow can be regarded as a dynamical system with multiple mobile actuators. In this case, one natural task is to investigate which formation of AVs could lead to a better performance of mixed traffic flow. Most related work focuses on understanding the potential of one single AV in mixed traffic flow~\cite{cui2017stabilizing,stern2018dissipation,wang2020controllability}, with a notable exception in~\cite{zheng2020smoothing}, where the case of multiple AVs is considered but the cooperative formation of AVs is not addressed.

{Another related topic is the \emph{actuator placement} or \emph{input selection} problem: identifying a subset of actuator placements from all possible choices to improve certain performance metrics.} This topic has been extensively discussed in a range of areas, such as mechanical systems~\cite{hiramoto2000optimal}, {power grids~\cite{qin2018submodularity}, 
and fluid dynamics~\cite{alfaro2015optimal}}, and typical metrics include controllability criteria~\cite{olshevsky2014minimal,summers2015submodularity} and robustness performance~\cite{summers2016actuator,munz2014sensor,zheng2020distributed}. To find the optimal actuator placement, existing research usually formulates a set function optimization problem that is NP-hard in general. Some theoretical results have been revealed for network systems with simple dynamics~\cite{olshevsky2014minimal,summers2015submodularity}. For efficient numerical computation, it is important to reveal some favorable properties such as submodularity~\cite{nemhauser1978an}, for which a simple greedy algorithm may return a near-optimal solution. 
To our best knowledge, the cooperative formation of AVs in mixed traffic flow, as well as the submodularity property, has not been discussed in the literature before. Existing formulations~\cite{hiramoto2000optimal,qin2018submodularity,summers2016actuator,munz2014sensor,alfaro2015optimal} or theoretical results~\cite{olshevsky2014minimal,summers2015submodularity} are not directly applicable, since the mixed traffic system has distinct and more complex dynamical properties.

\subsection{Contributions}

In this paper, we investigate the role of vehicular formation in improving traffic performance and then identify the optimal formation for AVs in mixed traffic. 
Motivated by the seminal experiments in~\cite{sugiyama2008traffic,stern2018dissipation}, we consider a single-lane ring-road setup which represents a simplified closed traffic system with no boundary conditions and also corresponds with a straight road of infinite length and periodic  dynamics~\cite{cui2017stabilizing, zheng2020smoothing, giammarino2020traffic}. We establish a set-function formulation to describe the mixed traffic performance, and this naturally leads to a set function optimization problem. We first consider the case of noncooperative formation under an independently designed ACC-type controller. Then, we further investigate the cooperative formation where the AV controllers are centrally  designed in different formations. 
Some initial discussions appeared in~\cite{li2020optimal}. Our main results of this paper are as follows.

\begin{enumerate}
	\item
	We introduce a set-function approach to describe different formations of AVs in mixed traffic. Our set-variable representation allows capturing the influence of both penetration rates and formations of AVs. Most previous work~\cite{van2006the,shladover2012impacts,talebpour2016influence,xie2019heterogeneous} only focuses on penetration rates of AVs based on numerical simulations, while a theoretical approach for analyzing the role of vehicular formation is lacking. Our optimization formulation based on the set-function approach fills such a gap and is able to quantify the optimal formation of AVs in mixed traffic.
	\item
	We discuss the case of noncooperative formation under ACC-type controllers, which is locally designed without cooperation between vehicles. A stability invariance property is revealed in the sense that different AV formations have no influence on the distribution of the closed-loop poles. Numerical experiments suggest the resulting $\mathcal{H}_2$ performance might be submodular. This result reveals a diminishing improvement property of traffic performance when increasing penetration rates of AVs under ACC strategies. Our results support and complement previous studies~\cite{van2006the,shladover2012impacts,talebpour2016influence} from a control-theoretic perspective.
	\item
	We then consider the case of cooperative formation, where the AVs' controllers are designed via centralized cooperative control. This strategy requires global traffic state information and quantifies the potential of a given formation of AVs in mitigating traffic perturbations~\cite{zheng2020smoothing,wang2020controllability}. We present explicit scenarios in which submodularity does not hold in this case. We further show that platooning of multiple AVs is not always the optimal formation. Interestingly, extensive numerical studies reveal two predominant optimal choices: \textit{platoon formation} and \textit{uniform distribution}. The optimal formation relies heavily on the string stability performance of HDVs. When HDVs have a poor string stability behavior, platoon formation might be the worst choice.
	\item
	We finally carry out nonlinear traffic experiments with a penetration rate of 20\% AVs. 
Results show that the platoon formation can achieve a satisfactory performance when traffic perturbation happens immediately ahead of the platoon. In other cases, however, distributing AVs uniformly achieves better performance in smoothing traffic flow. Together with the previous theoretical analysis, our results support the benefits of AVs in mitigating traffic perturbation and also suggest more opportunities for cooperative formation of multiple AVs beyond platooning. Mixed traffic systems can be more resilient to external disturbances by applying cooperative control to the AVs.
\end{enumerate}

The rest of this paper is organized as follows. Section \ref{Sec:Modeling} introduces the modeling process, and Section \ref{Sec:Formulation} presents the set function optimization formulation. Analysis on noncooperative formation and investigation on cooperative formation are presented in Section \ref{Sec:Pre-fixed} and Section \ref{Sec:Re-designed}, respectively. Section \ref{Sec:Results} demonstrates the numerical solutions of the optimal formation problem. Section \ref{Sec:Simulation} presents the results of nonlinear traffic simulation. We conclude the paper with extensive discussions in Section \ref{Sec:Conclusion}.

\section{Modeling Mixed Traffic Systems}

\label{Sec:Modeling}

\begin{figure}[t]
	\centering
	\subfigure[]
	{
		\includegraphics[scale=0.3]{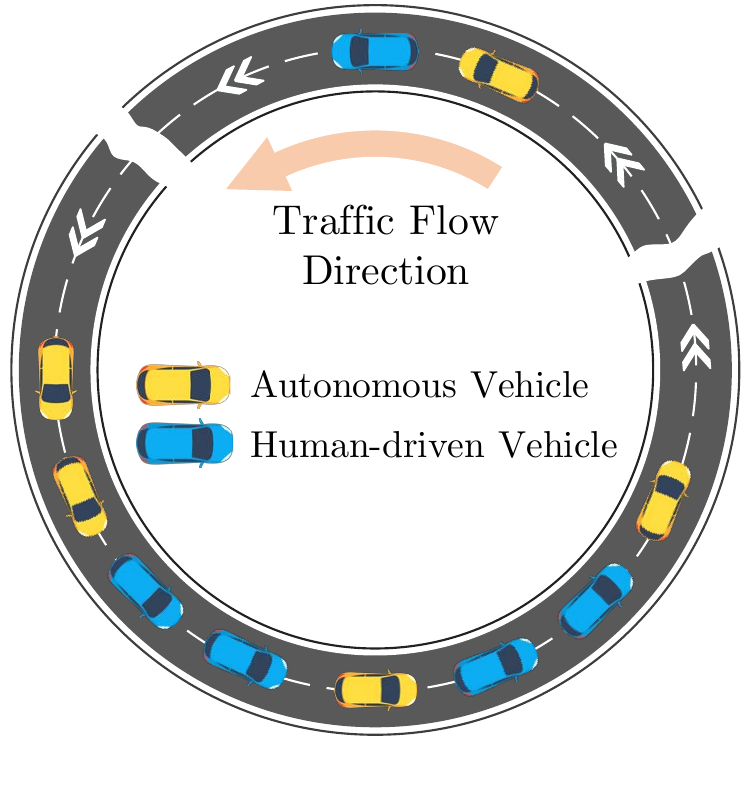}
	}
	\subfigure[]
	{
		\includegraphics[scale=0.3]{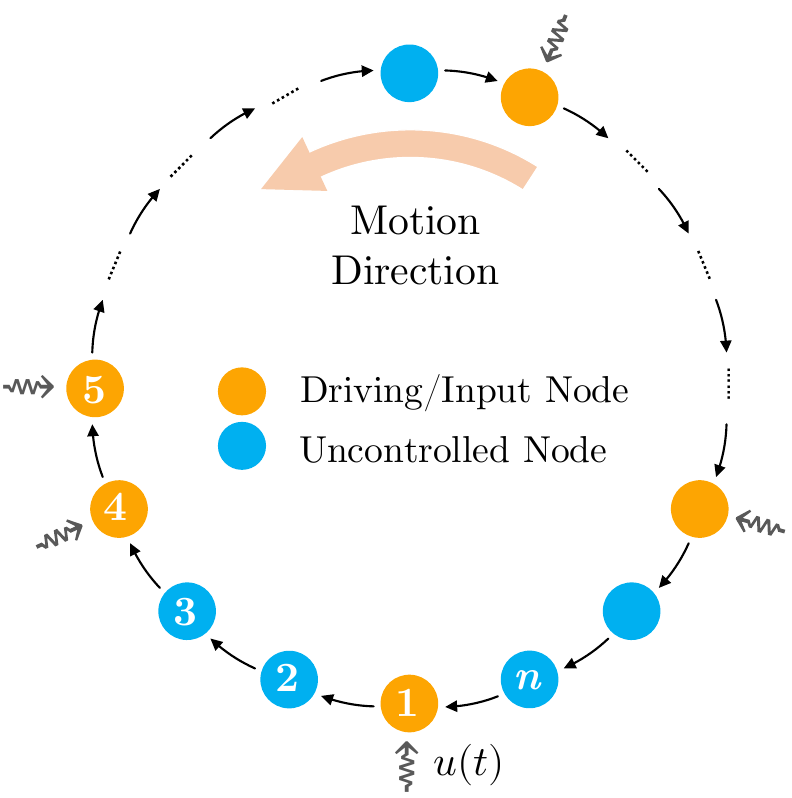}
	}
	\vspace{-2mm}
	\caption{Schematic for the ring-road traffic scenario. (a) The single-lane ring road scenario with AVs and HDVs. (b) A simplified network system schematic where AVs serve as driving/input nodes and HDVs are uncontrolled nodes. }
	\label{Fig:SystemModel}
	\vspace{-2mm}
\end{figure}

In this section, we present a dynamical model of mixed traffic systems in a ring-road setup. As shown in Fig.~\ref{Fig:SystemModel}, we consider a single-lane ring road of length  $ L $  with  $ n $  vehicles, among which there are  $ k $  AVs and  $ n-k $  HDVs. The vehicles are indexed from 1 to  $ n $ , and we define  $  \Omega = \{ 1,2, \ldots ,n \}$. Due to the single-lane setup, no lane changing behavior is considered, and we focus on the longitudinal dynamics.

The formation of AVs is characterized by their spatial location in mixed traffic, which is represented as a set variable
\begin{equation}
	S=\{i_1,\ldots,i_k \}\subseteq\Omega,
\end{equation}
with $i_1,\ldots,i_k$ denoting the spatial indices of AVs. Note that $\vert S \vert =k$, where $\vert \cdot \vert$ denotes the cardinality of a set. The position, velocity, and acceleration of vehicle $i$ are denoted as $p_i$, $v_i$ and $a_i$, respectively. The spacing of vehicle $i$, \ie, its relative (bumper-to-bumper) distance from vehicle $i-1$, is defined as $s_i=p_{i-1}-p_i$. The relative velocity is $\dot{s}_i=v_{i-1}-v_i$. The vehicle length is ignored without loss of generality.

According to typical HDV models, \eg, the optimal velocity model and the intelligent driver model, the longitudinal dynamics of an HDV can be described as~\cite{treiber2013traffic}
\begin{equation}\label{Eq:HDVNonlinearModel}
\dot{v}_i (t)=F(s_i (t),\dot{s}_i (t),v_i (t)),\;\;\; i\notin S,
\end{equation}
meaning that the acceleration of an HDV is determined by the relative distance, relative velocity and its own velocity. In an equilibrium traffic state, where $a_i=\dot{v}_i=0$ for $i\in \Omega$, each vehicle moves with the same equilibrium velocity $v^*$ and the corresponding equilibrium spacing $s^*$.
Based on~\eqref{Eq:HDVNonlinearModel}, the equilibrium state $(s^*,v^*)$ should satisfy
\begin{equation} \label{Eq:EqulibriumEquation}
	F(s^*,0,v^*) = 0.
\end{equation}

Assuming that each vehicle is under a small perturbation from $(s^*,v^*)$, we define the error state between actual and equilibrium state of vehicle $i$ as
\begin{equation}
\tilde{s}_i(t)=s_i(t)-s^*, \;
\tilde{v}_i(t)=v_i(t)-v^*.
\end{equation}
Applying the first-order Taylor expansion to~\eqref{Eq:HDVNonlinearModel}, a linearized model for each HDV is derived around the equilibrium state
\begin{equation}\label{Eq:LinearHDVModel}
\begin{cases}
\dot{\tilde{s}}_i(t)=\tilde{v}_{i-1}(t)-\tilde{v}_i(t),\\
\dot{\tilde{v}}_i(t)=\alpha_{1}\tilde{s}_i(t)-\alpha_{2}\tilde{v}_i(t)+\alpha_{3}\tilde{v}_{i-1}(t),\\
\end{cases} i\notin S,
\end{equation}
with $\alpha_1 = \frac{\partial F}{\partial s}, \alpha_2 = \frac{\partial F}{\partial \dot{s}} - \frac{\partial F}{\partial v}, \alpha_3 = \frac{\partial F}{\partial \dot{s}}$  evaluated at $s=s^*,v=v^*$. According to the real driving behavior, we have $\alpha _{1}>0$, $\alpha _{2}> \alpha _{3}>0$~\cite{jin2016optimal,cui2017stabilizing}. 
For each AV, the acceleration signal is directly used as the control input $u_i (t)$, and its car-following model is thus given by
\begin{equation}
\begin{cases}
\dot{\tilde{s}}_{i}(t)=\tilde{v}_{i-1}(t)-\tilde{v}_{i}(t),\\
\dot{\tilde{v}}_{i}(t)=u_{i}(t),
\end{cases} i \in S.
\end{equation}
Note that for the AVs, the spacing policy, \ie, the relationship between equilibrium spacing $s^*$ and equilibrium velocity $v^*$ can be manually set or follow the same equilibrium equation~\eqref{Eq:EqulibriumEquation} of the HDVs. 

To model traffic perturbations, we assume there exists a scalar disturbance signal $\omega_{i}(t)$ with finite energy in the acceleration of vehicle $i$ $(i \in\Omega)$. Lumping the error states of all the vehicles into one global state vector $x(t)=\left[\tilde{s}_{1}(t), \ldots, \tilde{s}_{n}(t), \tilde{v}_{1}(t), \ldots, \tilde{v}_{n}(t)\right]^{\tr}$ and letting $\omega(t)=[\omega_{1}(t)$ $, \ldots, \omega_{n}(t)]^{\tr}$, $u(t)=\left[u_{i_1}(t),\ldots,u_{i_k}(t)\right]^\tr$, the state-space model for the mixed traffic system is then written as
\begin{equation} \label{Eq:SystemModel}
\dot{x}(t)=A_{S} x(t)+B_{S} u(t)+H \omega(t),
\end{equation}
with
\begin{equation*}
\begin{aligned}
A_{S}&=\begin{bmatrix}{0} & {M_{1}} \\ {\alpha_{1}\left(I_{n}-D_{S}\right)} & {P_{S}}\end{bmatrix} \in \mathbb{R}^{2 n \times 2 n}, \\ B_{S}&=\begin{bmatrix}{\mathbb{e}_{i_{1}}, \mathbb{e}_{i_{2}}, \ldots, \mathbb{e}_{i_{k}}}\end{bmatrix} \in \mathbb{R}^{2 n \times k}, \\
H&=\begin{bmatrix}{0} \\ {I_{n}}\end{bmatrix}\in \mathbb{R}^{2 n \times n},
\end{aligned}
\end{equation*}
and
\begin{equation*}
\begin{aligned}
M_1 &= \begin{bmatrix}
-1 & & \cdots &1\\
1&-1& &\\
& \ddots & \ddots &\\
& & 1 &-1 	\end{bmatrix}\in \mathbb{R}^{n\times n},\\
D_S &= \mathrm{diag}\left(\delta_1,\delta_2,\ldots,\delta_n\right)\in \mathbb{R}^{n\times n},\\
P_S &= \begin{bmatrix}
-\alpha_2 \bar{\delta}_1 & & \cdots &\alpha_3 \bar{\delta}_1\\
\alpha_3 \bar{\delta}_2&-\alpha_2 \bar{\delta}_2& &\\
& \ddots & \ddots &\\
& & \alpha_3 \bar{\delta}_n &-\alpha_2 \bar{\delta}_n 	\end{bmatrix}\in \mathbb{R}^{n\times n}.
\end{aligned}
\end{equation*}
In this paper, we use $I_{n}$ and $\mathrm{diag}(\cdot)$ to denote an $n\times n$ identity matrix and a diagonal matrix, respectively. We also define a boolean variable $\delta_{i}$ to indicate whether vehicle $i$ is an AV, \ie,
\begin{equation} \label{Eq:Delta}
\delta_{i}=
\begin{cases}
0,\; \mathrm{if}\; i \notin S,\\
1,\; \mathrm{if}\; i \in S,
\end{cases}
\end{equation}
and let $\bar{\delta}_{i}=1-\delta_{i}$ indicate whether vehicle $i$ is an HDV. In the input matrix $B_{S}$, the vector $\mathbb{e}_{i_r}$ is a $2 n \times 1$ unit vector $(r=1,2, \ldots, k)$, with the $\left(i_{r}+n\right)$-th entry being one and the others being zeros.

It is clear that the control input $u_i (t)$  $(i\in S)$ of AVs plays an important role in the closed-loop performance of the mixed traffic system. {It is shown in~\cite{zheng2020smoothing,wang2020controllability} that the ring road mixed traffic system~\eqref{Eq:SystemModel} with one or more AVs ($k \geq 1$) is always stabilizable (but not controllable). This result guarantees the existence of stabilizing control input $u_i(t)$.}
In the following, we first analyze the case of noncooperative formation, where $u_i (t)$  $(i\in S)$ is designed individually. Specifically, we consider a noncooperative ACC-type controller that is similar to the HDVs' dynamics. 
Then, we consider the case of cooperative formation where the controllers of AVs are cooperatively designed based on centralized $\mathcal{H}_2$ optimal control. 
These are discussed in Sections~\ref{Sec:Pre-fixed} and~\ref{Sec:Re-designed}, respectively.

\begin{remark}
	The system matrices $A_S$ and $B_S$ in~\eqref{Eq:SystemModel} depend on the formation decision $S$, which is a set variable. This representation can not only describe the explicit spatial formation via its elements $S=\{i_1,\ldots,i_k \}$, but also the penetration rate, calculated by $\vert S \vert / \vert \Omega \vert=k/n$. Further, we show that this formulation allows for capturing the mixed traffic system performance naturally. Most existing work on mixed traffic flow focuses on the penetration rates only, usually described by a scalar index~\cite{van2006the,shladover2012impacts,talebpour2016influence,xie2019heterogeneous}; the role of different formations has not been discussed explicitly before. Note that a similar dynamical model was introduced in~\cite{zheng2020smoothing}, which is equivalent to~\eqref{Eq:SystemModel}, since the state vector in~\cite{zheng2020smoothing} can be transformed to $x(t)$ in~\eqref{Eq:SystemModel} by a permutation matrix. We choose the form~\eqref{Eq:SystemModel} due to its convenience to reflect the relationship between the system matrices $A_S$, $B_S$ and the formation decision $S$. Note that the ring-road scenario has a cyclic symmetry structure, 
	and this facilitates our analysis on the influence of different formations $S$ on mixed traffic performance. 
	Open straight roads are more practical scenarios, and dynamical models for mixed traffic in this case are also available~\cite{wang2020leading}.  \markend

\end{remark}

\section{Set Function Optimization Formulation}
\label{Sec:Formulation}

In this section, we first introduce a set-function formulation to describe the traffic system performance, and then describe a set function optimization approach to formulate the optimal formation problem.

\subsection{Set Function and Submodularity}

We now describe the performance of the mixed traffic system. Based on the dynamical model~\eqref{Eq:SystemModel}, we consider a general performance value function to measure the system-wide performance under formation $S$ of AVs
\begin{equation} \label{Eq:SetFunction}
J(S): 2^{\Omega} \rightarrow \mathbb{R}.
\end{equation}
Note that $J(S)$ is a set function, and we assume that a higher value of $J(S)$ indicates a better traffic performance.

Before presenting a precise choice of $J(S)$ in Section~\ref{Sec:H2Performance}, we introduce a notion of submodularity that plays a significant role in set function optimization~\cite{summers2015submodularity,nemhauser1978an}. Intuitively, submodularity describes a diminishing improvement property: adding an element to a smaller set gives a larger gain than adding it to a larger set. The formal definition is given below.

\begin{definition}[Submodularity~\cite{nemhauser1978an}] \label{De:Submodularity}
	A set function $f: 2^{\Omega} \rightarrow \mathbb{R}$ is called submodular if for all $A \subseteq B \subseteq \Omega$ and all elements $e \in \Omega$, it holds that
	\begin{equation} \label{Eq:SubmodularityDefinition}
	f(A \cup\{e\})-f(A) \geq f(B \cup\{e\})-f(B) .
	\end{equation}
\end{definition}

Submodularity plays an analogous role as concavity~in~discrete optimization~\cite{nemhauser1978an}. The following results are also  useful to check the submodularity of a set function .

\begin{definition}[Monotonicity~\cite{nemhauser1978an}] \label{De:Monotonicity}
	A set function $f: 2^{\Omega} \rightarrow \mathbb{R}$ is called non-increasing if for all $A \subseteq B \subseteq \Omega$, it holds that $f(A) \geq f(B)$.
\end{definition}

\begin{lemma} [\!\cite{nemhauser1978an}] \label{Lem:SubmodularityValidation}
A set function  $f: 2^{\Omega} \rightarrow \mathbb{R}$  is submodular if and only if the marginal improvement function  $ \Delta _{f} \left( e \vert  \cdot \right) :2^{ \Omega \setminus \{ e \} } \rightarrow R $, defined as
\begin{equation}
\Delta _{f} \left( e \vert  A \right) =f \left( A \cup  \{ e \}  \right) -f \left( A \right),
\end{equation}
are non-increasing for all  $ e \in  \Omega $.
\end{lemma}

\begin{figure}[t]
	\centering
	\includegraphics[scale=0.45]{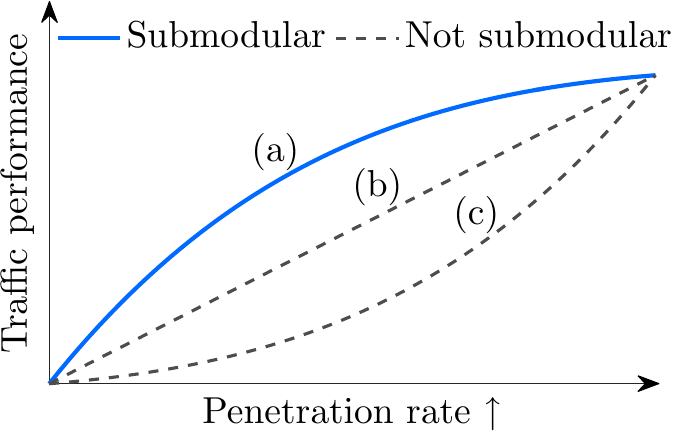}
	\vspace{-3mm}
	\caption{Interpretation of submodularity for traffic performance. (a) Submodular: the traffic performance is a concave function of the penetration rate, where the marginal improvement decreases as the penetration rate grows. (b) Modular: the marginal improvement remains constant, leading to a linear set function. (c) Supermodular: the marginal improvement increases as the penetration rate grows.}
	\label{Fig:SubmodularityIllustration}
	\vspace{-2mm}
\end{figure}

\begin{remark}
 Traffic performance typically improves as the penetration rate of AVs increases~\cite{van2006the,shladover2012impacts,talebpour2016influence}. If the performance metric $J(S)$ is submodular, then the marginal improvement brought by AVs diminishes as the increase of the penetration rate. This property leads to a concave and monotonically increasing curve of the traffic performance with respect to the penetrate rate; see Fig.~\ref{Fig:SubmodularityIllustration} for illustration. Unlike the formulations in~\cite{van2006the,shladover2012impacts,talebpour2016influence}, our set-function formulation~\eqref{Eq:SetFunction} contributes to a deeper understanding of  the influence of the penetration rates. \markend
\end{remark}

\subsection{$\mathcal{H}_2$ Performance for Mixed Traffic Performance}
\label{Sec:H2Performance}

To quantify the metric $ J \left( S \right)  $, controllability-related criteria have received significant attention; see \eg,~\cite{olshevsky2014minimal,summers2015submodularity}. However, it has been shown in~\cite{zheng2020smoothing,wang2020controllability} that a ring-road mixed traffic system is not completely controllable when  $  \vert S \vert  \geq 1 $; an uncontrollable mode associated with a zero eigenvalue always exists. Another typical metric for  $ J \left( S \right)  $  is the well-studied control-theoretic $ \mathcal{H}_{2} $  performance~\cite{summers2016actuator,munz2014sensor}.

\begin{definition} [$\mathcal{H}_2$ norm~\cite{skogestad2007multivariable}]
For a stable system $\dot{x}=Ax+H\omega$ with output $z=Cx$, the $\mathcal{H}_2$ norm of its transfer function $\mathbf{G}$ from $\omega$ to $z$ is defined as
\begin{equation}
\Vert \mathbf{G} \Vert _{2}=\sqrt{\mathrm{Tr} \left(  \int _{0}^{+\infty}Ce^{At}HH^{\tr}e^{A^{\tr}t}C^{\tr}dt \right) },
\end{equation}
\end{definition}
where $\mathrm{Tr}(\cdot)$ denotes the trace of a matrix.

\begin{lemma} [{\!\cite{skogestad2007multivariable}}]\label{Lem:H2Calculation}
For a stable system $\dot{x}=Ax+H\omega$ with output $z=Cx$, the $\mathcal{H}_2$ norm of its transfer function $\mathbf{G}$ from $\omega$ to $z$ can be computed by
\begin{equation}
\Vert \mathbf{G} \Vert_{2}^{2}=\!\inf _{X \succ 0}\left\{\mathrm{Tr}\left(C X C^{\tr}\right) | A X\!+\!X A^{\tr}+H H^{\tr} \!\preceq \!0\right\}.
\end{equation}
\end{lemma}

The $\mathcal{H}_{2}$ performance is able to capture the influence of traffic perturbations and the evolution of traffic waves incurred from the existence of traffic bottlenecks or the collective dynamics in drivers' behaviors; see Appendix~\ref{App:H2} for more discussions. In this paper, we consider the $\mathcal{H}_2$ performance as our main metric $J(S)$ to quantify the ability of different~formations.

\subsection{Optimal Formation Problem}

The potential of one single AV in stabilizing traffic flow and improving traffic performance has been demonstrated in~\cite{cui2017stabilizing,zheng2020smoothing,wang2020controllability}. In the case where multiple AVs coexist, the specific formation $S$ of AVs has a significant influence on the system-wide performance $J(S)$. 
We thus consider the following optimal formation problem. 
\begin{problem} \label{Pr:Main}
	Given $k$ AVs in the single-lane ring-road mixed traffic system~\eqref{Eq:SystemModel}, find an optimal spatial formation, \ie, $S=\left\{i_{1}, \ldots, i_{k}\right\} \subseteq \Omega,$ for the AVs to maximize the system-wide
	performance $J(S)$ for the entire traffic flow.
\end{problem}

\begin{figure}[t]
	\centering
	\subfigure[]
	{ \label{Fig:ExamplePlatoon}
		\includegraphics[scale=0.55]{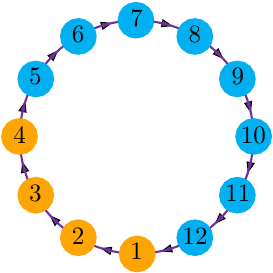}
	}
	\subfigure[]
	{\label{Fig:ExampleUniform}
		\includegraphics[scale=0.55]{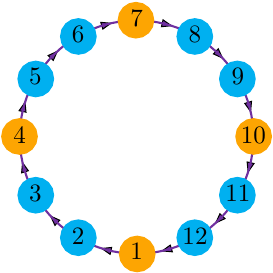}
	}
	\subfigure[]
	{
		\label{Fig:ExampleAbnormal}
		\includegraphics[scale=0.55]{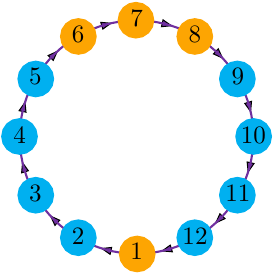}
	}
	\vspace{-2mm}
	\caption{Possible formations when $n=12,k=4$. Blue nodes: HDVs; yellow nodes: AVs. (a) Platoon formation ($S=\{1,2,3,4\}$). (b) Uniform distribution ($S=\{1,4,7,10\}$). (c) Abnormal formation ($S=\{1,6,7,8\}$).}
	\label{Fig:FormationExample}
	\vspace{-2mm}
\end{figure}

In Fig.~\ref{Fig:FormationExample}, we illustrate three examples of the formation of AVs in the ring-road mixed traffic system, when $n=12,k=4$ (the penetration rate is $33.3\%$). Possible formations include platoon formation (see Fig.~\ref{Fig:ExamplePlatoon}), uniform distribution (see Fig.~\ref{Fig:ExampleUniform}) and other abnormal cases (see Fig.~\ref{Fig:ExampleAbnormal}). We are interested in whether the prevailing platoon formation is the optimal choice for the mixed traffic scenario. Problem \ref{Pr:Main} can be formulated in an abstract way as follows.
\begin{equation} \label{Eq:ProblemFormulation}
\begin{aligned}
\max_S  \quad &J(S)\\
\mathrm{subject~to}\quad  & S \subseteq \Omega,|S|=k,
\end{aligned}
\end{equation}
where the optimal solution $S^*$ offers the optimal spatial formation for AVs in mixed traffic flow.

\begin{remark}
	Problem~\eqref{Eq:ProblemFormulation} is a standard set function optimization, which has been widely used in actuator placement; see \eg,~\cite{olshevsky2014minimal,summers2015submodularity,summers2016actuator}. For an LTI system given by $\dot{x}=Ax+Bu$, most existing results consider the case where the placement decision only affects the input matrix $B$~\cite{olshevsky2014minimal,summers2015submodularity,summers2016actuator}. In mixed traffic flow, however, AVs and HDVs have distinct dynamics. When we choose a different formation for AVs, the system matrix $A$ will also be changed. Therefore, in our system model~\eqref{Eq:SystemModel}, both the system matrix $A_S$ and the input matrix $B_S$ rely on the formation $S$ of AVs, and the results in~\cite{olshevsky2014minimal, summers2015submodularity, summers2016actuator} are not applicable. Note that the set-function approach can deal with the actuator placement problem based on various types of dynamics, including the linear-time invariant system model or the partial differential equation model that is common in fluid dynamics~\cite{alfaro2015optimal}. It is also  interesting  to address the cooperative formation from a macroscopic traffic perspective~\cite{treiber2013traffic}. \markend
\end{remark}



\section{Analysis Under Noncooperative Formation}
\label{Sec:Pre-fixed}

The proposed modeling approach, including the dynamics model~\eqref{Eq:SystemModel} and the set function formulation~\eqref{Eq:ProblemFormulation}, where the formation of AVs is represented as a set variable, allows us to reveal useful properties of the mixed traffic system. 
We here analyze the case of noncooperative formation where AVs adopt an ACC-type controller. In particular, we focus on the closed-loop stability of the mixed traffic system and the submodularity of the corresponding $\mathcal{H}_2$ performance.

\subsection{Stability Invariance}
\label{Sec:StabilityInvariance}

\begin{figure*}[t]
	\centering
	\includegraphics[width=.85\textwidth]{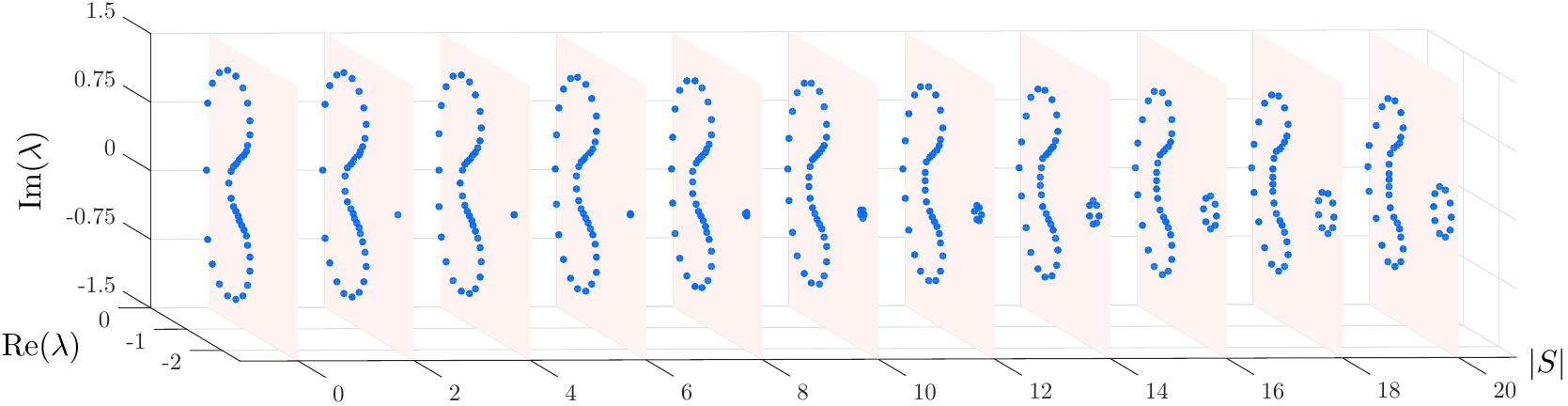}
	\vspace{-2mm}
	\caption{Illustration of the stability invariance property: the distribution of the closed-loop poles $\lambda$ of the mixed traffic system when AVs adopt an ACC-type strategy ($n=20$). The distribution is independent to the formation $S$ at a fixed value of $|S|$. In the linearized HDV model~\eqref{Eq:LinearHDVModel}, $\alpha_{1}=0.94$, $\alpha_{2}=1.5$, $\alpha_{3}=0.9$; in the ACC controller~\eqref{Eq:ACC}, $k_{s}=0.1$, $k_{v}=1 $.}
	\label{Fig:EigenvalueDistribution}
	\vspace{-2mm}
\end{figure*}

Similar to the driving behavior~\eqref{Eq:HDVNonlinearModel}, an ACC-equipped AV usually utilizes local information, such as relative distance and velocity to the preceding vehicle, to adjust its velocity and maintain a pre-specified spacing~\cite{li2017dynamical}. Motivated by~\cite{cui2017stabilizing,xie2019heterogeneous}, we consider a modified ACC strategy for each AV ($i\in S$)
\begin{equation} \label{Eq:ACC}
u_{i}(t)=\left(\alpha_{1}-k_{s}\right) \tilde{s}_{i}(t)-\left(\alpha_{2}+k_{v}\right) \tilde{v}_{i}(t)+\alpha_{3} \tilde{v}_{i-1}(t),
\end{equation}
which is augmented from the linearized HDV model~\eqref{Eq:LinearHDVModel}, with $k_s,k_v$ being two constant feedback gains. This ACC strategy~\eqref{Eq:ACC} is similar to the HDV's linearized dynamics, facilitating our subsequent analysis.
The feedback gain $k_s,k_v$ is assumed to remain unchanged under different formations, \ie, no cooperative controller design is used. Accordingly, we call the system under this ACC-type controller~\eqref{Eq:ACC} as a \textit{noncooperative formation}.

Substituting controller~\eqref{Eq:ACC} into the mixed traffic system~\eqref{Eq:SystemModel}, the closed-loop model for the traffic system becomes
\begin{equation} \label{Eq:ACCSystemModel}
\dot{x}(t)=\widehat{A}_{S} x(t)+H \omega(t),
\end{equation}
where
\begin{equation*}
	\widehat{A}_{S}=\begin{bmatrix}0 & M_{1} \\ \alpha_{1} I_{n}-k_{s} D_{S} & M_{2}-k_{v} D_{S}\end{bmatrix},
\end{equation*}
with
\begin{equation*}
M_{2}=\begin{bmatrix}-\alpha_{2} & & \cdots & \alpha_{3} \\ \alpha_{3} & -\alpha_{2} & & \\ & \ddots & \ddots & \\ & & \alpha_{3} & -\alpha_{2}\end{bmatrix}.
\end{equation*}
Based on the closed-loop model~\eqref{Eq:ACCSystemModel}, we consider the stability property under different formations $S$. 
A linear time-invariant (LTI) system $\dot{x}=A x$ is asymptotically stable, if and only if all the eigenvalues of $A$ have negative real parts. As shown
in~\cite{zheng2020smoothing,wang2020controllability}, the mixed traffic system~\eqref{Eq:SystemModel} always has a zero eigenvalue, whose algebraic multiplicity is one. Therefore, the mixed traffic system~\eqref{Eq:SystemModel} is not asymptotically stable, but can be made Lyapunov stable using stabilizing controllers~\cite{zheng2020smoothing}. In this paper, stability refers to the Lyapunov sense; see Appendix~\ref{App:ProofStability} for a precise definition of Lyapunov stability. 

One first analytical result is a \emph{stability invariance} property of the ring-road mixed traffic system.

\begin{theorem} \label{Th:Stability}
	Consider a linearized ring-road mixed traffic system with $k$ AVs and $n-k$ HDVs given by~\eqref{Eq:ACCSystemModel}, where AVs adopt a noncooperative ACC controller~\eqref{Eq:ACC}. Then, the distribution of the closed-loop poles of 	$\widehat{A}_{S}$ in~\eqref{Eq:ACCSystemModel} is independent of the formation $S$ of AVs.
\end{theorem}
 
 The proof is postponed to Appendix~\ref{App:ProofStability}.
This result reveals that the mixed traffic system~\eqref{Eq:ACCSystemModel} under different formations of AVs has the same distribution of closed-loop poles when the number of AVs is fixed. Fig.~\ref{Fig:EigenvalueDistribution} illustrates the distribution of the closed-loop poles of~\eqref{Eq:ACCSystemModel} under a typical parameter setup~\cite{jin2016optimal} when $n=20$, which remains the same under different formations $S$ when $|S|$ is fixed. The closed-loop stability of mixed traffic systems has received significant attention in previous 
studies, but most of them focus on the impact of penetration rates~\cite{talebpour2016influence,xie2019heterogeneous}. Interestingly, Theorem~\ref{Th:Stability} presents the stability invariance property of the ring-road mixed traffic system at a fixed value of the penetration rate $|S| /|\Omega|,$ when AVs follow a noncooperative ACC-type strategy~\eqref{Eq:ACC}.

\subsection{Submodularity of $\mathcal{H}_2$ Performance}
\label{Sec:Submodularity_ACC}

In addition to stability, the closed-loop traffic system should have a good ability to dissipate disturbances. To quantify this, we proceed to consider the $\mathcal{H}_2$ performance, as discussed in Section~\ref{Sec:H2Performance}. The output of the closed-loop mixed traffic system~\eqref{Eq:ACCSystemModel} is defined as
\begin{equation} \label{Eq:Output_ACC}
	z_{1}(t)=\begin{bmatrix}Q_{1}^{1 / 2} & 0 \\ 0 & Q_{2}^{1 / 2}\end{bmatrix} x(t),
\end{equation}
where $Q_{1}=\mathrm{diag}\left(\gamma_{s}, \ldots, \gamma_{s}\right), Q_{2}=\mathrm{diag}\left(\gamma_{v}, \ldots, \gamma_{v}\right)$. The
weight coefficients $\gamma_{s}, \gamma_{v}>0$ represent the penalty for
spacing error and velocity error, respectively. The $\mathcal{H}_{2}$ norm
of the transfer function $\mathbf{G}_{1}(S)$ from disturbance $\omega$ to output
$z_{1}$ is used to describe the influence of perturbations on system~\eqref{Eq:ACCSystemModel}. 

We then calculate the performance value function $J(S)$ in~\eqref{Eq:SetFunction} as follows, denoted as $J_{1}(S)$.
\begin{equation} \label{Eq:PerformanceFunction_ACC}
J_{1}(S):=-\Vert \mathbf{G}_{1}(S)\Vert_{2}^{2}.
\end{equation}
The negative sign is used for normalization, and a larger performance value function represents a better traffic performance.
To characterize the submodularity of $J_{1}(S),$ we first need to investigate the monotonicity of the marginal improvement $\Delta_{J_{1}}(e | S)$ for all $e \in \Omega$ according to Lemma~\ref{Lem:SubmodularityValidation}. Thanks to the circulant structure of our ring-road setup, it is sufficient to examine the monotonicity of $\Delta_{J_{1}}(1 | S)$.

Since the $\mathcal{H}_{2}$ norm needs to be evaluated numerically via Lemma~\ref{Lem:H2Calculation}, it is non-trivial to obtain an analytical expression of $J_{1}(S)$, and so is $\Delta_{J_{1}}(1 | S)$. We thus examine the submodularity of $J_{1}(S)$ by exploiting a numerical algorithm based on Lemma~\ref{Lem:SubmodularityValidation} (see Algorithm \ref{Alg:Submodularity} in Appendix~\ref{App:Submodularity}). The main idea is to generate a series of random sequences of the marginal improvement $\left\{\Delta_{J_{1}}\left(1 | S_{i}\right)\right\}$, $i=1,2, \ldots, n,$ where
\begin{equation} \label{Eq:RandomSequence}
	\vert S_{i}\vert=i, \, S_{i} \subseteq S_{i+1},\, i=1, \ldots, n-1 ;\; S_{n}=\Omega.
\end{equation}
Given a set of parameter values and a sufficiently large number of experiments, if all random sequences $\left\{\Delta_{J_{1}}\left(1 | S_{i}\right)\right\}$ are non-increasing, we can make a reasonable conjecture that $J_{1}(S)$ is submodular under the parameter setup according to Lemma~\ref{Lem:SubmodularityValidation}. If one counterexample is found, we can conclude that $J_{1}(S)$ is not submodular.

\begin{table}[tb]
	\begin{center}
		\caption{Parameter Setups in Fig.~\ref{Fig:Submodularity_ACC}}\label{Tb:Submodularity_ParameterSetup}
		\vspace{-1mm}
		\begin{tabular}{cccccc}
\toprule
			&$\alpha_1$ & $\alpha_2$ & $\alpha_3$  & $k_s$ & $k_v$ \\ \hline
			(a)&0.94 & 1.5 & 0.9  & 0.1 & 1\\
			(b)&0.94 & 1.5 & 0.9  & 0.3 & 3\\
			(c)&0.5 & 2.5 & 0.5  & 0.1 & 1\\
			(d)&0.5 & 2.5 & 0.5  & 0.3 & 3\\
			\bottomrule
		\end{tabular}
	\end{center}
\vspace{-2mm}
\end{table}

\begin{figure}[t]
		\setlength{\abovecaptionskip}{0em}
    \setlength{\belowcaptionskip}{0em}
	\centering
	\subfigure[]
	{ \includegraphics[scale=0.42]{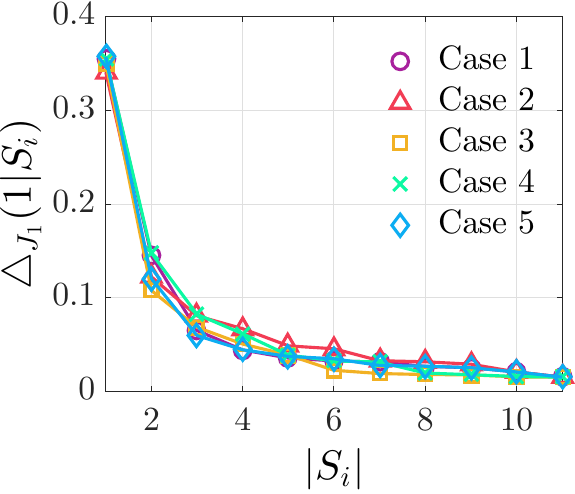}
	}
	\subfigure[]
	{
	\includegraphics[scale=0.42]{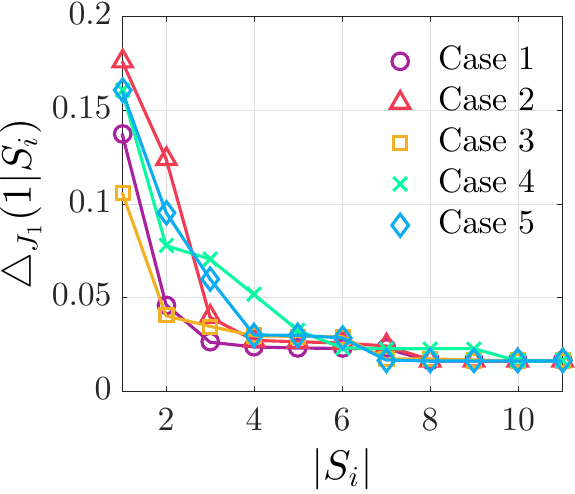}
	}
\subfigure[]
	{ \includegraphics[scale=0.42]{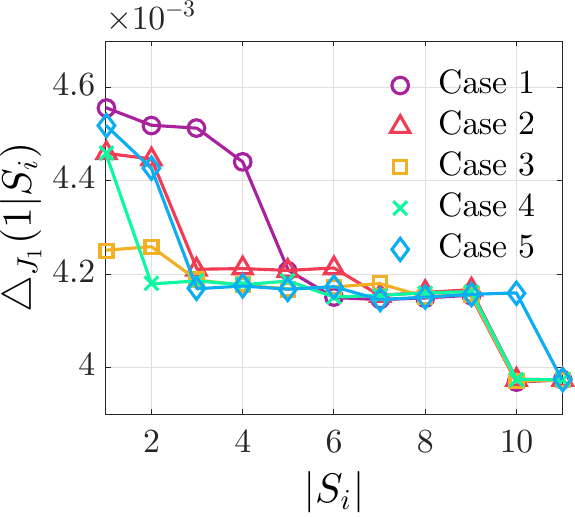}
}
\subfigure[]
{
	\includegraphics[scale=0.42]{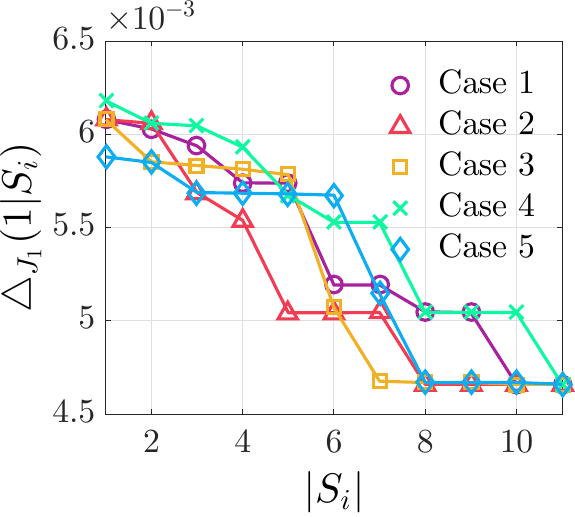}
}
	\caption{Five random results of the marginal improvement sequence
		$\{\Delta_{J_{1}}\left(1 | S_{i}\right)\}$ where $n=12, \gamma_{s}=0.01, \gamma_{v}=0.05$. Parameter values
		are shown in Table~\ref{Tb:Submodularity_ParameterSetup} with corresponding orders.}
	\label{Fig:Submodularity_ACC}
\end{figure}

We consider the case where $n=12,$ and utilize Algorithm~\ref{Alg:Submodularity} to carry out $200$ random experiments for four different parameter setups; see Table~\ref{Tb:Submodularity_ParameterSetup}. The setup of $\alpha_{1}, \alpha_{2}, \alpha_{3}$ represents two kinds of HDV driving behaviors~\cite{orosz2010traffic}: string unstable for (a)(b) and string stable for (c)(d), while the setup of $k_s,k_v$ represents different degrees of improvement in string stability brought by AVs. From all the random experiments, we observe that the sequences $\left\{\Delta_{J_{1}}\left(1 | S_{i}\right)\right\}$ are always non-increasing under each parameter setup. Five random cases under each setup are illustrated in Fig.~\ref{Fig:Submodularity_ACC}. Based on these numerical results, we conjecture the following result.

\begin{conjecture} \label{Conj:ACC_Submodular}
	The $\mathcal{H}_2$ performance $J_1 (S)$ defined in~\eqref{Eq:PerformanceFunction_ACC} under the noncooperative ACC controller is submodular.
\end{conjecture}

Our extensive numerical experiments suggest the conjecture above holds. A theoretical proof is interesting but technically difficult, and we leave it for future work.

\begin{remark}
	A wide range of studies analyzed the influence of ACC strategies on traffic flow~\cite{van2006the,shladover2012impacts,talebpour2016influence}, and most of them have shown by traffic simulations that the traffic performance improves as the penetration rate of ACC-equipped vehicles increases. Here, we make a further step and observe that the $\mathcal{H}_{2}$ performance of mixed traffic might be submodular.
	Our results support that only a few AVs can dramatically improve traffic dynamics and smooth traffic flow~\cite{stern2018dissipation,zheng2020smoothing}, but also indicate that the marginal performance improvement diminishes when the penetration rate of ACC-equipped vehicles increases. \markend
\end{remark}

	It is worth noting that the ACC-type controllers are independently designed and noncooperative (see~\eqref{Eq:ACC}). Their parameter setup typically remains fixed in case of any formation change of AVs. Ideally, the control strategy of AVs should be \emph{redesigned according to their current formation in mixed traffic flow}. The specific controllers of AVs might be distinct from each other in different formations. In the following, we seek to redesign the controllers of AVs for different formations in a centralized way, 
	which quantifies the potential of each cooperative formation in dissipating traffic perturbations.

\section{Analysis Under Cooperative Formation}

\label{Sec:Re-designed}

In this section, we consider the case of cooperative formation where the controllers of AVs are redesigned for different formations in a centralized optimal way, and the resulting $\mathcal{H}_{2}$ performance is chosen as the explicit form of the performance value function $J(S)$ in~\eqref{Eq:SetFunction}. 

\subsection{Cooperative Controller}
\label{Sec:OptimalController}

We assume that all the involved vehicles have vehicle-to-vehicle (V2V) or vehicle-to-infrastructure (V2I) communication capabilities, and the global state of the entire mixed traffic system is available to the AVs for designing their input. Given a formation $S$, we consider static state feedback control as
\begin{equation} \label{Eq:CooperativeController}
u=-K_{S} x, \;K_{S} \in \mathbb{R}^{k \times 2n}.
\end{equation}
%
We use $z_2(t)$ to denote a performance output for the global mixed traffic system
\begin{equation} \label{Eq:Output_Optimal}
z_2(t) = \begin{bmatrix} Q^{\frac{1}{2}} \\0 \end{bmatrix}x(t) +  \begin{bmatrix} 0 \\R^{\frac{1}{2}} \end{bmatrix}u(t),
\end{equation}
where $Q^{\frac{1}{2}} = \mathrm{diag}\left(\gamma_s,\ldots,\gamma_s,\gamma_v,\ldots,\gamma_v\right)\in \mathbb{R}^{2n \times 2n} $ and $R^{\frac{1}{2}} = \mathrm{diag}(\gamma_u,\ldots,\gamma_u)\in \mathbb{R}^{k \times k} $. The weight coefficients $\gamma_s,\gamma_v,\gamma_u>0$ represent the penalty for spacing error, velocity error and control input, respectively. When applying the controller $u=-K_S x$, the dynamics of the closed-loop mixed traffic system then become
\begin{equation}
\begin{aligned}
\dot{x}(t) &= (A_S - B_S K_S)x(t) + H \omega (t), \\
z_2(t) &= \begin{bmatrix} Q^{\frac{1}{2}} \\ -R^{\frac{1}{2}}K_S \end{bmatrix}x(t).
\end{aligned}
\end{equation}

The $\mathcal{H}_{2}$ norm of the transfer function $\mathbf{G}_{2}(S)$ from disturbance $\omega$ to output $z_{2}$ is utilized to describe the influence of perturbations on the traffic system for a given formation decision $S$. Then, the optimal cooperative feedback gain $K_{S}$ of AVs can be obtained by solving
\begin{equation} \label{Eq:H2Control}
\min_{K_S} \; \lVert \mathbf{G}_2 (S) \rVert_2^2,
\end{equation}
which is a standard $\mathcal{H}_2$ optimal control problem~\cite{skogestad2007multivariable}. We present brief steps to obtain a convex reformulation for~\eqref{Eq:H2Control}.

Using Lemma~\ref{Lem:H2Calculation} and a standard variable substitution $K_S=Z X^{-1}$, problem~\eqref{Eq:H2Control} can be equivalently converted to
\begin{equation*} 
\begin{aligned}
\min _{X, Z}\;\;&\Vert \mathbf{G}_{2}(S)\Vert_{2}^{2}=\mathrm{Tr}(Q X)+\mathrm{Tr}\left(R Z X^{-1} Z^{\tr}\right)\\
\mathrm{subject~to}\;&\left(A_{S} X-B_{S} Z\right)+\left(A_{S} X-B_{S} Z\right)^{\tr}+H H^{\tr} \preceq 0,\\
 & X \succ 0 .
\end{aligned}
	\end{equation*}
Using the Schur complement and introducing $Y \succeq Z X^{-1} Z^{\tr}$, the problem above can be reformulated as the  convex optimization problem~\cite{skogestad2007multivariable}
\begin{align}
\min_{X,Y,Z} \; \;& \Vert \mathbf{G}_{2}(S)\Vert_{2}^{2}={\mathrm{Tr}}(QX)+{\mathrm{Tr}}\left(RY\right) \nonumber\\
\mathrm{subjec to} \; \;& (A_S X-B_S Z)+(A_S X-B_S Z)^{\tr} + HH^{\tr} \!\preceq \!0, \nonumber \\
&\begin{bmatrix}
Y & Z \\ Z^\tr & X\end{bmatrix}\succeq 0,\; X \succ 0.  \label{Eq:LMIOptimalControl}
\end{align}
Problem~\eqref{Eq:LMIOptimalControl} can be further converted into a standard semidefinite program, which can be solved efficiently via existing solvers, \eg, Mosek~\cite{mosek2010mosek}. After obtaining the optimal solution $(X,Y,Z)$ of Problem~\eqref{Eq:LMIOptimalControl}, the feedback gain in the controller~\eqref{Eq:CooperativeController} can be recovered by
	$K_S=ZX^{-1}$.

\begin{remark}
	The cooperative feedback gain~\eqref{Eq:CooperativeController} depends on the explicit choice of the formation, indicating that $K_{S}$ is redesigned in different formations $S$. In~\eqref{Eq:CooperativeController}, it is assumed that the global traffic state information~\eqref{Eq:SystemModel} is available to AVs via V2V/V2I communication. This assumption allows to achieve the optimal $\mathcal{H}_2$ performance, revealing the potential of AVs in improving traffic performance under a given formation $S$. For practical implementation, however, the AVs might be able to only acquire partial information from neighboring vehicles. 
	In this case, the optimal cooperative controller has a structural constraint, which becomes nontrivial to compute; see the discussions on structured optimal control in~\cite{wang2020controllability,jovanovic2016controller}. 
	In our work, the optimal formation is identified based on access to state information of all the vehicles in traffic flow, characterized by~\eqref{Eq:CooperativeController} and~\eqref{Eq:LMIOptimalControl}, but the case with practical communication constraints is worth further investigation in future work. Another potential approach for addressing communication constraints is to incorporate macroscopic traffic information, similarly to traffic flow prediction and control~\cite{abadi2014traffic}. It is an interesting topic to reveal the optimal formation with macroscopic information available.
\end{remark}

\subsection{Submodularity of $\mathcal{H}_2$ Performance}

Given a formation $S$ of AVs, the optimal feedback gain $K_{S}$ can be obtained by solving~\eqref{Eq:LMIOptimalControl}. Meanwhile, the optimal value of $\min _{K_{S}}\left\|\mathbf{G}_{2}(S)\right\|_{2}^{2}$ indicates the minimum influence of perturbations on the entire traffic flow. Accordingly, the expression of the performance value function $J(S)$ in~\eqref{Eq:SetFunction} can be given by
\begin{equation} \label{Eq:Performance_Optimal}
J_2(S):=-\min_{K_S}  \lVert \mathbf{G}_2 (S) \rVert_2^2,
\end{equation}
which is denoted as $J_{2}(S)$ in the following discussion. 
The negative sign is used for normalization.

Based on this new reformulation~\eqref{Eq:Performance_Optimal} of the performance value function, we observe that submodularity does not hold for $J_2 (S)$; a simple counterexample is shown as follows. 

\begin{example}
Assume $\alpha_{1}=0.5, \alpha_{2}=2.5, \alpha_{3}=0.5$ and $\gamma_{s}=0.01, \gamma_{v}=0.05, \gamma_{u}=0.1$. Let $S_{1}=\{4,9,10\}$ and $S_{2}=\{2,3,4,9,10\}$, which implies $S_1 \subseteq S_2 $. Then we can compute directly that
$$
\begin{aligned} J_2\left(S_{1} \cup\{1\}\right)=-0.5982,\, J_2\left(S_{1}\right)=-0.5003; \\
J_2\left(S_{2} \cup\{1\}\right)=-0.7860,\, J_2\left(S_{2}\right)=-0.6910. \end{aligned}
$$
It is clear to see that
$$
\begin{aligned}
&J_2\left(S_{1} \cup\{1\}\right)-J_2\left(S_{1}\right) =-0.098 \\
 \leq \;\;&J_2\left(S_{2} \cup\{1\}\right)-J_2\left(S_{2}\right)=-0.095, \end{aligned}
$$
which violates~\eqref{Eq:SubmodularityDefinition} in Definition \ref{De:Submodularity}, indicating that $J_2(S)$ is not submodular. 
\end{example}
\begin{fact} \label{Fact:Submodularity}
	The $\mathcal{H}_{2}$ performance $J_{2}(S)$ defined in~\eqref{Eq:Performance_Optimal} under the cooperative controller from~\eqref{Eq:H2Control} is not a submodular function. 
\end{fact}

\begin{figure}[t]
	\setlength{\abovecaptionskip}{0em}
    \setlength{\belowcaptionskip}{0em}
	\centering
	\subfigure[]
	{ \includegraphics[scale=0.42]{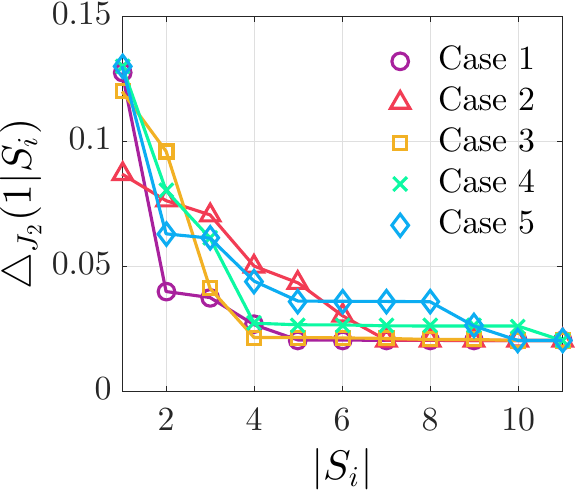}
	}
	\subfigure[]
	{
		\includegraphics[scale=0.42]{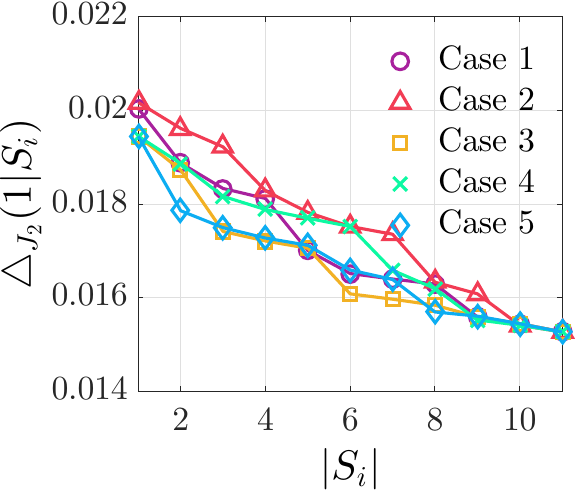}
	}
	\caption{Five random results of $\left\{\Delta_{J_{2}}\left(1 | S_{i}\right)\right\}$ where $n=12, \gamma_{s}=0.01, \gamma_{v}=0.05, \gamma_{u}=1 \times 10^{-6}$. (a) $\alpha_{1}=0.94, \alpha_{2}=1.5, \alpha_{3}=0.9$. (b) $\alpha_{1}=0.5, \alpha_{2}=2.5, \alpha_{3}=0.5$.}
	\label{Fig:Submodularity_Optimal}
\end{figure}

\begin{remark}
	Note that one difference between the performance output $z_{2}(t)$ in~\eqref{Eq:Output_Optimal} and that in~\eqref{Eq:Output_ACC} is the existence of the penalty $\gamma_{u}$ for the control input $u(t)$, which constrains the control energy. Then, the dimension of the control input increases as the growth of $|S|$. We can also investigate whether there exist certain conditions where $J_{2}(S)$ is submodular. In particular, we consider the case where the penalty $\gamma_{u}$ is sufficiently small compared with $\gamma_{s}$, $\gamma_{v}$. We let $\gamma_{s}=0.01, \gamma_{v}=0.05, \gamma_{u}=1 \times 10^{-6}$. This case indicates that the control objective mainly aims to minimize the state error of each vehicle under the perturbation. Since the semidefinite program in~\eqref{Eq:LMIOptimalControl} can only be solved numerically, it is nontrivial to obtain the analytical expression of $J_{2}(S)$. Therefore, similarly to Section~\ref{Sec:Submodularity_ACC}, Algorithm \ref{Alg:Submodularity} is again utilized to examine the monotonicity of $\Delta_{J_{2}}(1 | S)$. In the case where $n=12$, $200$ random experiments were conducted for two different parameter setups. Five random results are shown in Fig.~\ref{Fig:Submodularity_Optimal}. We observe that $\{\Delta_{J_{2}}\left(1 | S_{i}\right)\}$ are always non-increasing sequences under each random case, indicating the function might be submodular under this condition. \markend
\end{remark}

\subsection{{Reformulation of Optimal Formation}}

As shown in Section~\ref{Sec:OptimalController}, the cooperative controller derived from the optimization problem~\eqref{Eq:LMIOptimalControl} offers an optimal feedback gain $K_{S}$ that achieves the best performance in minimizing the influence of the perturbations under a given formation $S$. This bound reveals to which extent the AVs given by $S$ can improve the traffic flow. 
We then reformulate the original optimal formation problem~\eqref{Eq:ProblemFormulation} to address Problem~\ref{Pr:Main}, given as
\begin{equation} \label{Eq:H2ProblemFormulation}
\begin{aligned}
\max_{S}  \quad &J_2(S)=-\min_{K_S}  \lVert \mathbf{G}_2 (S) \rVert_2^2\\
\mathrm{subject~to}\quad  &S \subseteq \Omega,|S|=k.
\end{aligned}
\end{equation}
In~\eqref{Eq:H2ProblemFormulation}, the inner optimization problem~\eqref{Eq:LMIOptimalControl} needs to be first solved to calculate the value of $J_{2}(S)$ for a given formation decision $S$. Since it is proved in \cite{zheng2020smoothing} that the mixed traffic system with one or more AVs is always stabilizable, there exist stabilizing feedback gains $K_{S}$ under which the $\mathcal{H}_{2}$ norm of $\mathbf{G}_{2}(S)$ is finite, when $|S| \geq 1$. This guarantees the existence of a finite value of
$ \min _{K_{S}}\Vert \mathbf{G}_{2}(S)\Vert_{2}^{2} $.

Note that submodularity not only captures a diminishing improvement property, but also plays a critical role in solving set-function optimization problems. Specifically, for maximizing a submodular and monotone increasing set function, a simple greedy algorithm can return a near-optimal solution~\cite{summers2015submodularity,nemhauser1978an}. However, as shown in Fact~\ref{Fact:Submodularity}, $J_2 (S)$ defined in~\eqref{Eq:Performance_Optimal} is not submodular in general. 
Hence, the greedy algorithm in previous work, \eg,~\cite{summers2015submodularity}, cannot provide guarantees when solving Problem~\eqref{Eq:H2ProblemFormulation}. Since our main focus is to find out the \emph{exact optimal formation} of AVs, as described in Problem~\ref{Pr:Main}, the \emph{true optimal solution} to Problem~\eqref{Eq:H2ProblemFormulation} needs to be identified. Therefore, based on the proposed formulation~\eqref{Eq:H2ProblemFormulation}, the brute force method, \ie, enumerating all possible subsets of cardinality $k$, is a straightforward approach to obtain the true optimal formation solution.



\section{Numerical Studies on Optimal Formation}

\label{Sec:Results}

In this section, we present extensive numerical studies on the optimal formation of AVs in mixed traffic flow based on formulation~\eqref{Eq:H2ProblemFormulation}.

\begin{figure}[t]
	\centering
	\subfigure[]
	{ \label{Fig:OVMSpacing}
		\includegraphics[scale=0.41]{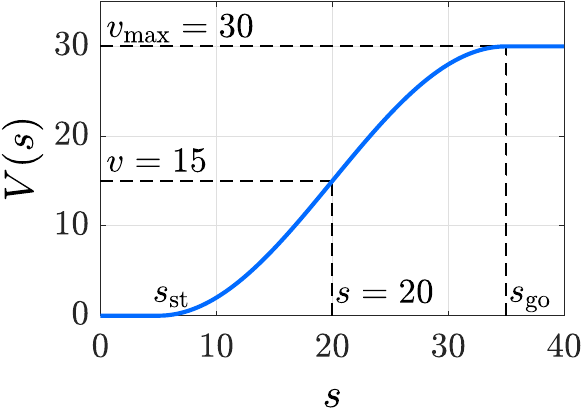}
	}
	\subfigure[]
	{\label{Fig:OVMVelocityDot}
		\includegraphics[scale=0.41]{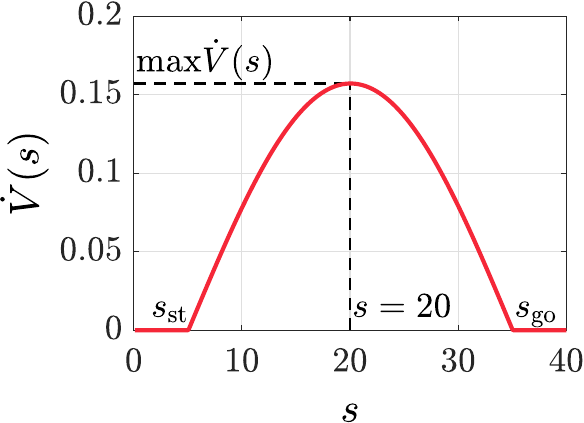}
	}
	\vspace{-2mm}
	\caption{Typical profile of the spacing-dependent desired velocity $V(s)$ and its derivative $ \dot{V}(s)$ when $\alpha = 0.6,\beta = 0.9, v_{\max} =30,s_{\text{st}}=5,s_{\text{go}}=35$.}
	\label{Fig:OVMPolicy}
	\vspace{-2mm}
\end{figure}

\subsection{Numerical Setup}

To clarify the physical interpretation of parameter setups, we utilize an explicit car-following model, the optimal velocity model (OVM)~\cite{jin2016optimal,orosz2010traffic}, in our numerical studies. In OVM, we denote $\alpha, \beta>0$ as the driver's sensitivity coefficients to the difference between current and desired velocity and the relative velocity between the preceding and ego vehicle, respectively. The specific model of HDVs~\eqref{Eq:HDVNonlinearModel} under OVM is~\cite{jin2016optimal}
\begin{equation}   \label{Eq:OVM}
F(\cdot)=\alpha \left(V(s_{i}(t))-v_{i}(t)\right)+\beta
\dot{s}_{i}(t),
\end{equation}
where $V(\cdot)$ denotes the spacing-dependent desired velocity, typically given by
\begin{equation}
\label{Eq:OVM_DesiredVelocity}
V(s)=\begin{cases}
0, &s\le s_{\mathrm{st}};\\
f_v(s), &s_{\mathrm{st}}<s<s_{\mathrm{go}};\\
v_{\max}, &s\ge s_{\mathrm{go}},
\end{cases}
\end{equation}
with
\begin{equation}
\label{Eq:OVM_SpacingPolicy}
f_{v}(s)=\frac{v_{\max }}{2}\left(1-\cos (\pi
\frac{s-s_{\text{st}}}{s_{\text{go}}-s_{\text{st}}})\right).
\end{equation}
In the OVM model, the coefficients in the linearized HDV model~\eqref{Eq:LinearHDVModel} can be calculated as
\begin{equation}
\alpha_{1}=\alpha \dot{V}\left(s^{*}\right), \alpha_{2}=\alpha+\beta, \alpha_{3}=\beta,
\end{equation}
where $\dot{V}\left(s^{*}\right)$ denotes the derivative of $V(\cdot)$ at equilibrium spacing $s^*$. Fig.~\ref{Fig:OVMPolicy} illustrates the curves of $V(s)$ and $\dot{V}(s)$ under a typical parameter setup as that in~\cite{jin2016optimal}.

\subsection{Case Studies and Two Predominant Formations}
\label{Sec:Result1}

\begin{figure}[t]
	\centering
	\subfigure[$\alpha=1.4, \beta=1.8, s^{*}=10$]
	{ \label{Fig:FormationResult_Platoon}
		\includegraphics[scale=0.55]{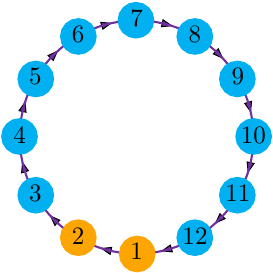}
		\includegraphics[scale=0.55]{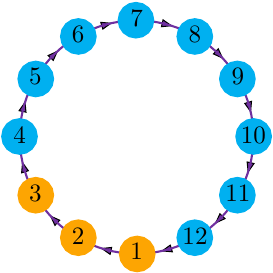}
		\includegraphics[scale=0.55]{Figs/Fig9_1c-eps-converted-to.pdf}
	}
	\subfigure[$\alpha=0.6, \beta=0.9, s^{*}=20$]
	{\label{Fig:FormationResult_Uniform}
		\includegraphics[scale=0.55]{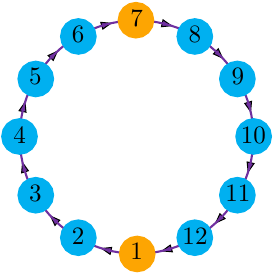}
		\includegraphics[scale=0.55]{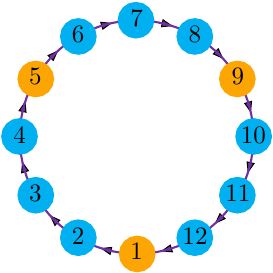}
		\includegraphics[scale=0.55]{Figs/Fig9_2c-eps-converted-to.pdf}
	}
	\subfigure[$\alpha=0.9, \beta=1.3, s^{*}=16$]
	{
		\label{Fig:FormationResult_Abnormal}
		\includegraphics[scale=0.55]{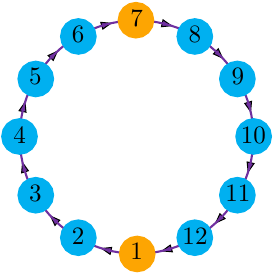}
		\includegraphics[scale=0.55]{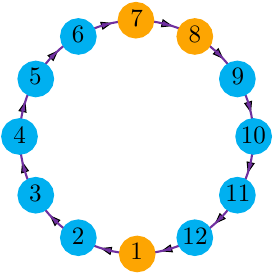}
		\includegraphics[scale=0.55]{Figs/Fig9_3c-eps-converted-to.pdf}
	}
	\vspace{-2mm}
	\caption{Optimal formation under specific cases $(n=12)$. In each panel, $k=2,3,4$ from left to right. $v_{\max }=30, s_{\mathrm{st}}=5, s_{\mathrm{go}}=35, \gamma_{s}=	0.01, \gamma_{v}=0.05, \gamma_{u}=0.1$.}
	\label{Fig:FormationResult_CaseStudy}
	\vspace{-2mm}
\end{figure}

\begin{figure*}[tb]
	\centering
	\subfigure[$n=12,k=2$]
	{ \label{Fig:FormationResult1}
		\includegraphics[scale=0.4]{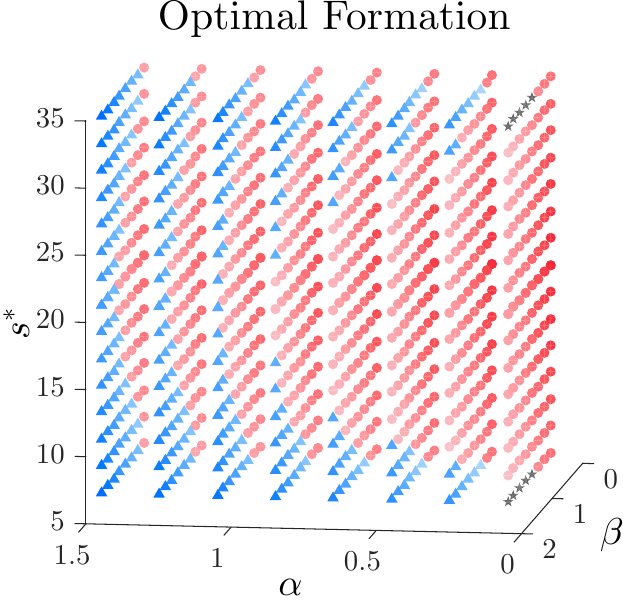}
		\includegraphics[scale=0.4]{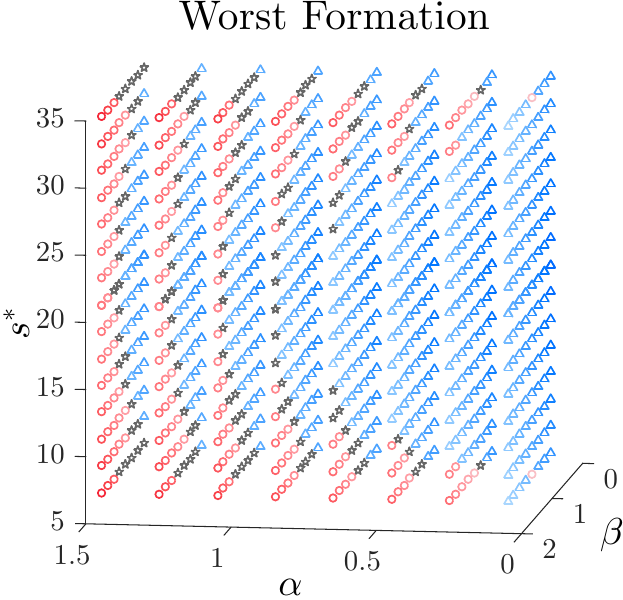}
	}
	\subfigure[$n=12,k=4$]
	{ \label{Fig:FormationResult2}
		\includegraphics[scale=0.4]{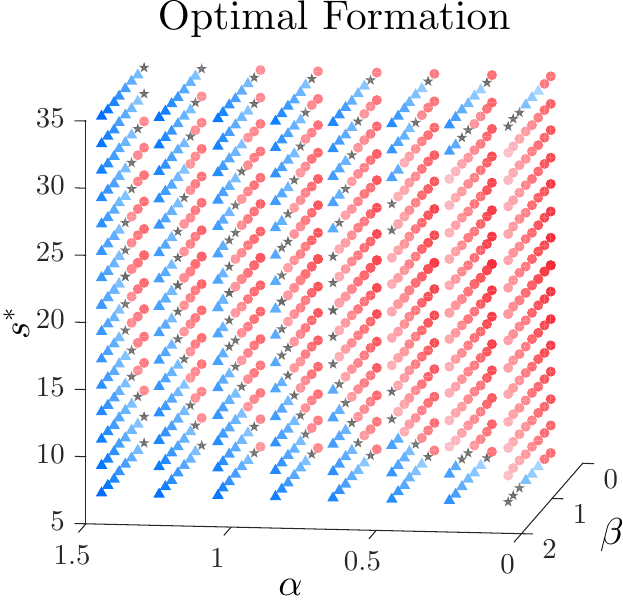}
		\includegraphics[scale=0.4]{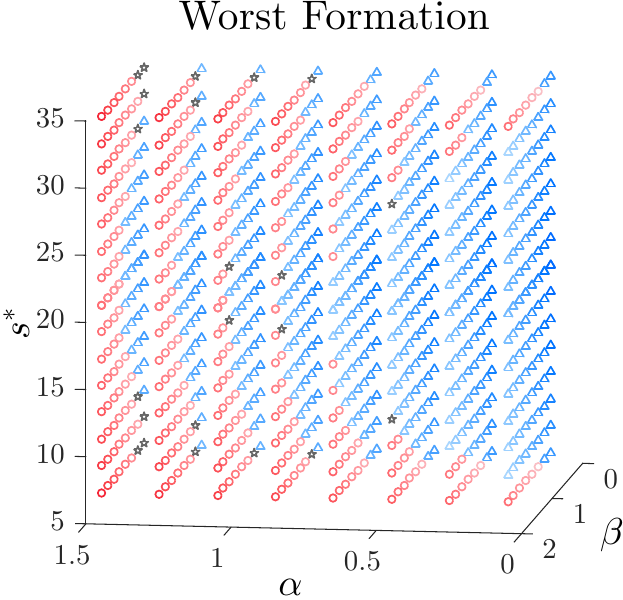}
	}
	\subfigure[$n=12,k=2$]
	{ \label{Fig:FormationResult3}
		\includegraphics[scale=0.4]{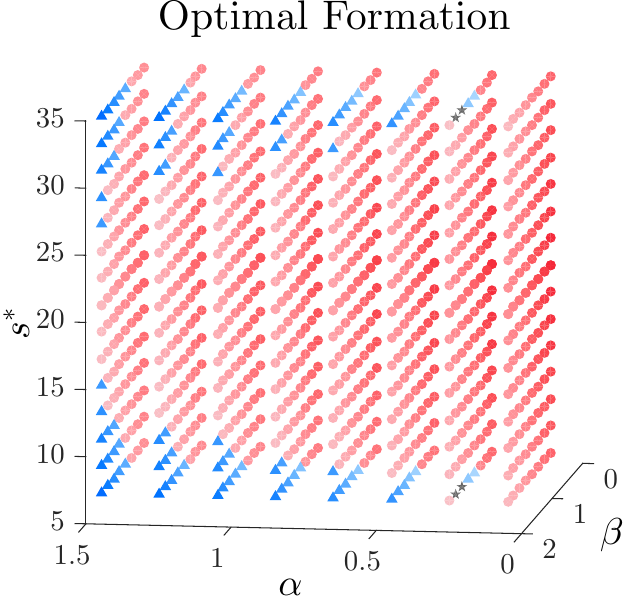}
		\includegraphics[scale=0.4]{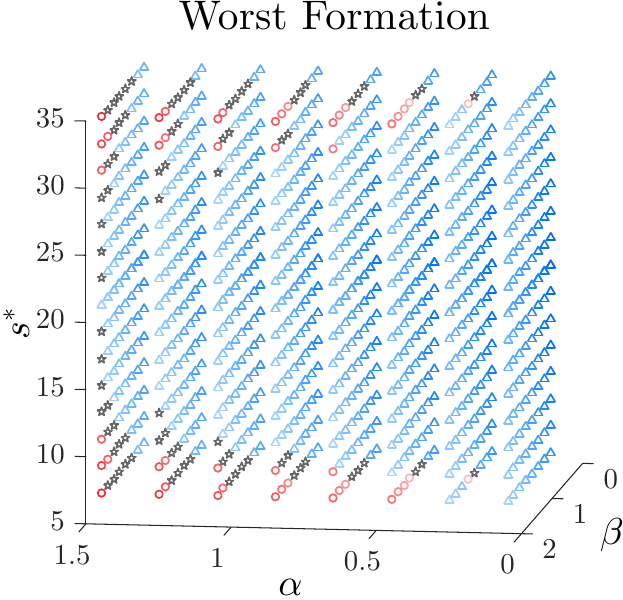}
	}
	\subfigure[$n=12,k=4$]
	{ \label{Fig:FormationResult4}
		\includegraphics[scale=0.4]{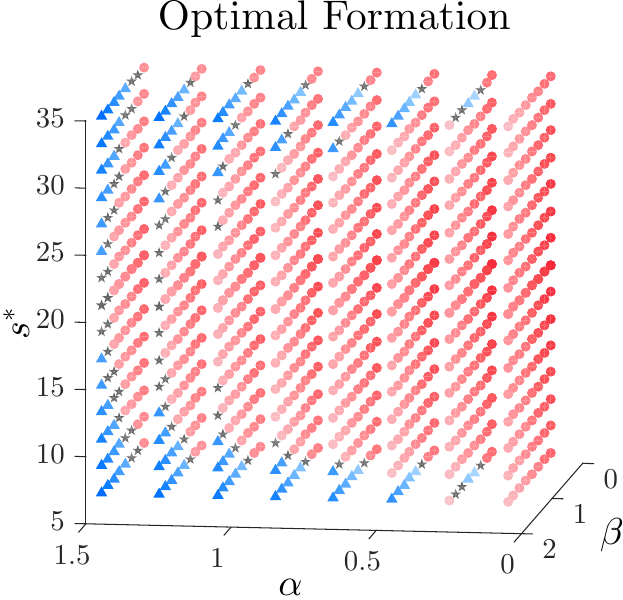}
		\includegraphics[scale=0.4]{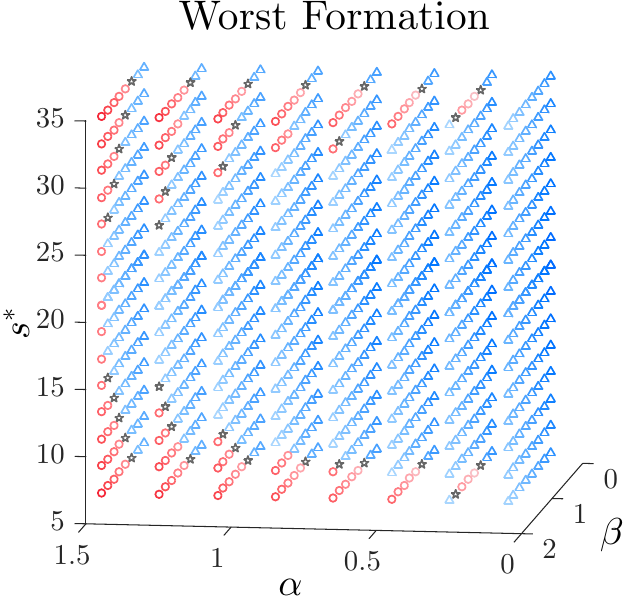}
	}
	\vspace{-3mm}
	\caption{Optimal and worst formation at various parameter setups. Red circles, blue triangles, and gray stars denote uniform distribution, platoon formation, and abnormal formations, respectively. In each panel, the left figure shows the optimal formation, where the darker the red, the larger the value of $\xi$; the darker the blue, the smaller the value of $\xi$. In contrast, the right figure shows the worst formation, where the darker the blue, the larger the value of $\xi$; the darker the red, the smaller the value of $\xi$. (a)(b) $\gamma_s=0.01,\gamma_v=0.05,\gamma_u=0.1$. (c)(d) $\gamma_s=0.03,\gamma_v=0.15,\gamma_u=0.1$.}
	\label{Fig:FormationResult}
	\vspace{-3mm}
\end{figure*}

Our first numerical study focuses on several specific cases of parameter setups to address Problem~\ref{Pr:Main}, \ie, identify the optimal formation of AVs in mixed traffic flow. We fix $v_{\max }=30, s_{\mathrm{st}}=5, s_{\mathrm{go}}=35 \text { and let } \gamma_{s}=0.01, \gamma_{v}=0.05, \gamma_{u}=0.1$. Then we observe that the numerical solution of the optimal formation relies on the parameter setup in the OVM model, \ie, the car-following behavior of HDVs. Three specific parameter setups are considered and their corresponding optimal formations are shown in Fig.~\ref{Fig:FormationResult_CaseStudy} when $n=12$, $k=2,3,4 $. Platoon formation, uniform distribution or certain abnormal formations could be the optimal formation. Note that the abnormal formations shown in Fig.~\ref{Fig:FormationResult_Abnormal} can be viewed as a transition pattern between platoon formation and uniform distribution. They can be regarded as a uniform distribution of several mini platoons, which also received research attention in the literature~\cite{zheng2016stabilitymargin}.

We then proceed to solve~\eqref{Eq:H2ProblemFormulation} in various parameter setups. The number of vehicles is set to $n=12$, $k=2$ or $4$, corresponding to a penetration rate of $16.7 \%$ or $33.3 \%$, respectively. The other parameters are fixed as $v_{\max }=30$, $ s_{\mathrm{st}}=5, s_{\mathrm{go}}=35$, and we discretize the three key parameters $\alpha, \beta, s^{*}$ within a common range~\cite{orosz2010traffic}: $\alpha \in$ $[0.1,1.5], \beta \in[0.1,1.5], s^{*} \in[5,35]$. Two different setups
of the weight coefficients $\gamma_{s}, \gamma_{v}, \gamma_{u}$ in the performance output~\eqref{Eq:Output_Optimal} are considered, representing a balanced penalty for the state perturbation of the mixed traffic system and the control input of the AVs~\cite{zheng2020smoothing,wang2020controllability}. Via solving~\eqref{Eq:H2ProblemFormulation}, the worst formation can be also identified.

The numerical results of optimal formation and worst formation are illustrated in Fig.~\ref{Fig:FormationResult}. As we clearly  observe, there exist two predominant patterns for optimal formations: platoon formation and uniform distribution, which are represented by blue triangles and red circles, respectively. This result holds regardless of the specific number $k$ of the AVs or the value of weight coefficients in~\eqref{Eq:Output_Optimal}. Along the boundary, there exist some abnormal formation patterns, represented by gray stars. Interestingly, we observe that the optimal formation and the worst formation have an evident relationship: when uniform distribution is optimal, platoon formation usually becomes the worst, and vice versa. This result indicates that the prevailing platoon formation might be the optimal formation, but could also be the worst choice, depending on the parameter setup of HDV models, \ie, the human driving behavior.

We further consider four other sets of weight coefficients. The values and their corresponding proportion of the optimal formation is listed in Tables~\ref{Tb:OptimalFormation_1} and~\ref{Tb:OptimalFormation_2}. Still, uniform distribution and platoon formation remain the two predominant patterns in all setups. In particular, when fixing $\gamma_u$ (the penalty for control inputs of the AVs) and increasing $\gamma_s$ and $\gamma_v$ (the penalty for the error state of the mixed traffic system), the proportion of uniform distribution being the optimal formation clearly grows up. This reveals that uniform distribution is more likely to be the optimal formation when emphasizing more on mitigating the state error of the mixed traffic system. By contrast, when constraining the control input of the AVs, which may contribute to better fuel economy and driving comfort of the AVs, platoon formation tend to be the optimal choice.

\begin{table}[t]
	\begin{center}
		\caption{Proportion of being the Optimal Formation at Different Weight Coefficients when $n=12,k=2$}\label{Tb:OptimalFormation_1}
		\vspace{-1mm}
		\begin{threeparttable}
		\setlength{\tabcolsep}{3mm}{
		\begin{tabular}{ccc|ccc}
\toprule
			$\gamma_s$ & $\gamma_v$ & $\gamma_u$ & UD & PF & AF \\ \hline
			0.001 & 0.005 &0.1 & 37\% & 62\% & 1\%\\
			0.01 & 0.05 &0.1 & 71\% & 28\% & 1\%\\
			0.03 & 0.15 &0.1 & 89\% & 10\% & 1\%\\
			0.05 & 0.25 &0.1 & 93\% & 6\% & 1\%\\
			\bottomrule
		\end{tabular}}
		\begin{tablenotes}
		\footnotesize
		\item[1] UD: uniform distribution; PF: platoon formation; AF: abnormal formation.
		\end{tablenotes}
		\end{threeparttable}
	\end{center}
	\vspace{-5mm}
\end{table}

\begin{table}[tb]
	\begin{center}
		\caption{Proportion of being the Optimal Formation at Different Weight Coefficients when $n=12,k=4$}\label{Tb:OptimalFormation_2}
		\vspace{-1mm}
		\setlength{\tabcolsep}{3mm}{
		\begin{tabular}{ccc|ccc}
\toprule
			$\gamma_s$ & $\gamma_v$ & $\gamma_u$ & UD & PF & AF\\ \hline
			0.001 & 0.005 &0.1 & 32\% & 65\% & 3\%\\
			0.01 & 0.05 &0.1 & 57\% & 33\% & 10\%\\
			0.03 & 0.15 &0.1 & 82\% & 11\% & 7\%\\
			0.05 & 0.25 &0.1 & 90\% & 6\% & 4\%\\
			\bottomrule
		\end{tabular}}
	\end{center}
	\vspace{-4mm}
\end{table}

\subsection{Poor HDV Car-following Behavior Requires Formation of AVs beyond Platooning}
\label{Sec:Result2}

We further investigate the explicit relationship between the optimal formation and the HDV parameter setup. It is observed that the string stability performance of HDVs' car-following behavior has a strong impact on the optimal formation of AVs in mixed traffic. A string of multiple vehicles is called string unstable if the amplitude of certain oscillations, \eg, spacing error or velocity error, are amplified along the propagation upstream the traffic flow~\cite{orosz2010traffic}. As shown in~\cite{orosz2010traffic}, the condition for strict string stability of OVM after linearization is
\begin{equation}
\alpha+2\beta \geq 2 \dot{V} (s^*).
\end{equation}
Here we define a string stability index $\xi$ as
\begin{equation}
\xi:=\alpha+2 \beta-2 \dot{V}\left(s^{*}\right).
\end{equation}
Note that a larger value of $\xi$ indicates a better string stability
behavior. In our parameter setup, $\dot{V}\left(s^{*}\right)$ decreases as $\vert s^*-$
$20 \vert$ grows up, as shown in Fig.~\ref{Fig:OVMVelocityDot}. Therefore, larger values of
$\alpha$, $\beta$ or $\vert s^*-20 \vert $ lead to a larger value of $\xi$, \ie, a better string stability performance of HDVs.

In Fig.~\ref{Fig:FormationResult}, we also utilize the color darkness to indicate the value of $\xi$. 
Then the relationship between string stability of HDVs and the optimal formation of AVs can be clearly observed. At a larger value of $\xi$ (in lower left and upper left of each panel), platoon formation appears to be the optimal choice. In contrast, when $\xi$ is small (in middle right of each panel), indicating a poor string stability behavior of HDVs, uniform distribution achieves the best performance while platoon formation becomes the worst.

Note that most HDVs tend to have a poor string stability behavior due to drivers' large reaction time and limited perception abilities~\cite{sugiyama2008traffic,orosz2010traffic}. This result indicates that platoon formation might limit the potential of AVs to improve real-world traffic performance, compared to other possible formations in the mixed traffic environment. When HDVs have a poor string stability performance, distributing AVs uniformly allows AVs to maximize their capabilities in suppressing traffic instabilities and mitigating undesired perturbations. Instead, when all human drivers have better driving abilities, organizing all the AVs into a platoon appears to be a better choice.

\begin{figure}[tb]
	\centering
	\subfigure[]
	{ \label{Fig:SystemScaleComparison1}
		\includegraphics[scale=0.43]{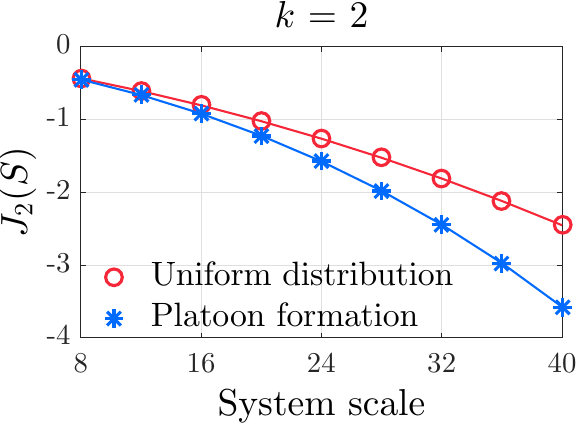}
		\includegraphics[scale=0.43]{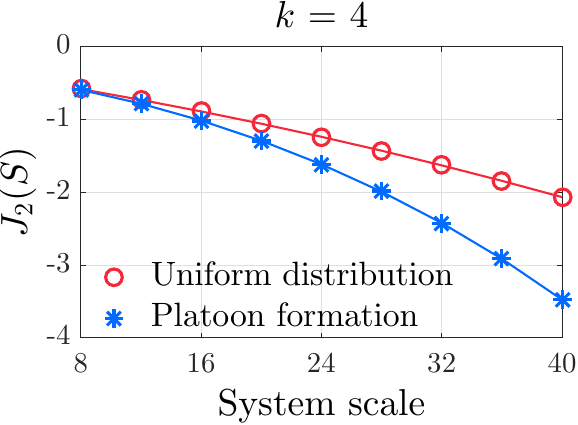}
	}
	\subfigure[]
	{\label{Fig:SystemScaleComparison2}
		\includegraphics[scale=0.43]{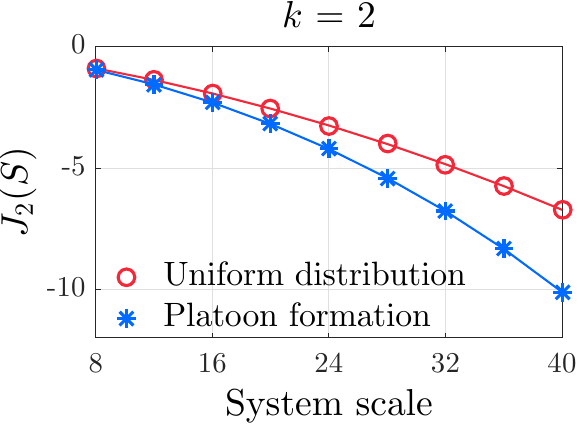}
		\includegraphics[scale=0.43]{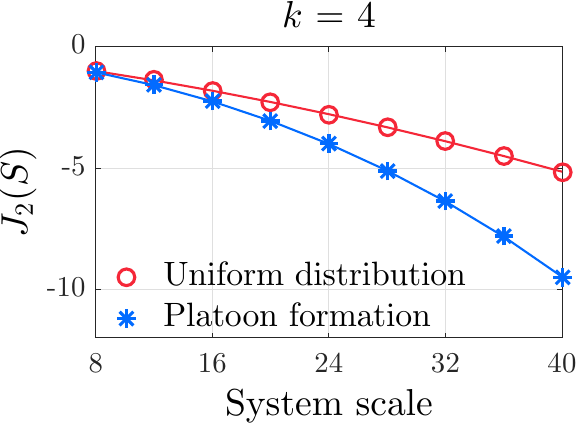}
	}
	\vspace{-3mm}
	\caption{Comparison of the performance value function between platoon formation and uniform distribution at different system scales. In OVM model,  $\alpha = 0.6,\beta = 0.9, s^*=20, v_{\max} =30,s_{\text{st}}=5,s_{\text{go}}=35$. (a) $\gamma_s=0.01,\gamma_v=0.05,\gamma_u=0.1$. (b) $\gamma_s=0.03,\gamma_v=0.15,\gamma_u=0.1$.}
	\label{Fig:SystemScaleComparison}
\end{figure}

\subsection{Comparison Between Platoon Formation and Uniform Distribution}

We carry out another numerical study to make further comparisons between the two predominant formations at different system scales $n \in [8,40]$. In Sections~\ref{Sec:Result1} and~\ref{Sec:Result2}, we consider different OVM parameter setups, and focus on the case where $n=12$. Here we vary the system scale, and fix the OVM model to a typical setup for human's driving behavior as that in~\cite{jin2016optimal}. The comparison of the performance value function $J_2(S)$ under these two formations is demonstrated in Fig.~\ref{Fig:SystemScaleComparison} ($k=2$ or $4$). Recall that a larger value of $J_2(S)$ denotes a better performance, \ie, a smaller influence of the perturbations on the entire traffic system. It is observed that in this typical parameter setup of human drivers, uniform distribution is optimal to~\eqref{Eq:ProblemFormulation}, while platoon formation is the worst. Moreover, as shown in Fig.~\ref{Fig:SystemScaleComparison}, the performance gap between the two formations is rapidly enlarged as the system scale grows up. This result indicates that at a large system scale, \ie, a low penetration rate of AVs, there could exist a huge performance difference between platoon formation and other possible formations, \eg, uniform distribution. In the near future when we only have a few AVs on the road, platooning might not be the optimal choice for improving the traffic performance.

\section{Nonlinear Traffic Simulation}
\label{Sec:Simulation}
Finally, this section presents numerical results from nonlinear traffic simulations at a large system scale to compare the performance of the two predominant formations revealed before: uniform distribution and platoon formation.

We consider a single-lane ring road with circumference $L=800\, \mathrm{m}$ containing $40$ vehicles, where there are eight AVs, \ie, $|S|=k=8$. The penetration rate of AVs is $20 \%$ in this setup. For the uniform distribution, we let $S=\{3,8,13,18,23,28,33,38\}$, while for the platoon formation, we assume $S=\{17,18,19,21,21,22,23,24\}$. The nonlinear OVM model~\eqref{Eq:OVM} is utilized to describe the car-following behavior of HDVs. To reflect the real-world traffic environment, we consider a heterogeneous parameter setup for each HDV motivated by the configuration in~\cite{jin2016optimal}: $\alpha=0.6+\mathbb{U}[-0.1,0.1]$,  $\beta=0.9+\mathbb{U}[-0.1,0.1]$, $s_{\mathrm{go}}=35+\mathbb{U}[-5,5]$, where $\mathbb{U}[\cdot]$ denotes the uniform distribution for parameter perturbation. Other parameters are set as $v_{\max}=30$, $s_{\mathrm{st}}=5$, $v^*=15$, and the equilibrium spacing $s^*$ of each HDV can be calculated according to~\eqref{Eq:EqulibriumEquation}. Upon this parameter setup, in the equilibrium traffic state, each vehicle has an equilibrium velocity of $15\,\mathrm{m/s}$ and an average equilibrium spacing of $20\,\mathrm{m}$.

\begin{figure}[t]
	\centering
	\subfigure[]
	{
		\includegraphics[scale=0.37]{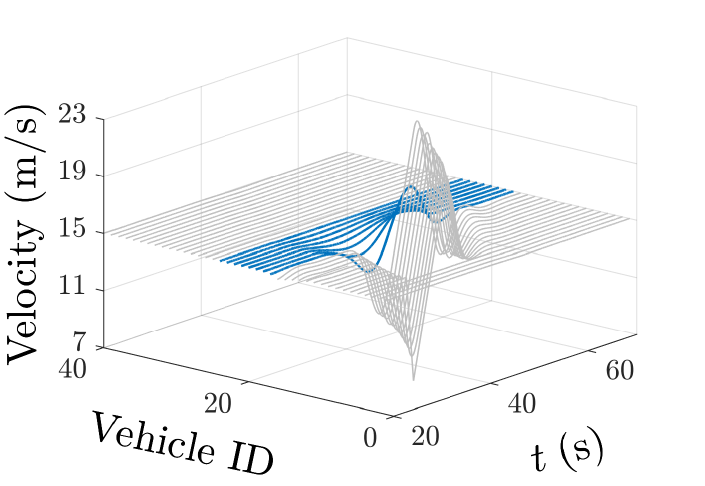}
		\includegraphics[scale=0.35]{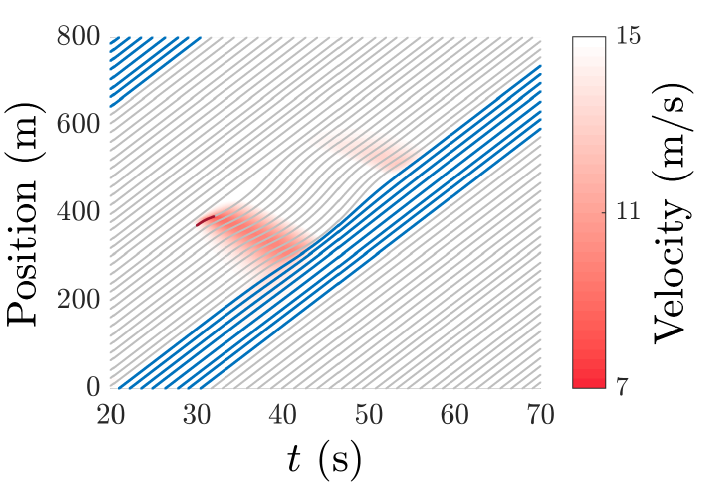}
		\label{Fig:Simulation_Trajectory_Platoon}
	}
		\subfigure[]
	{
		\includegraphics[scale=0.37]{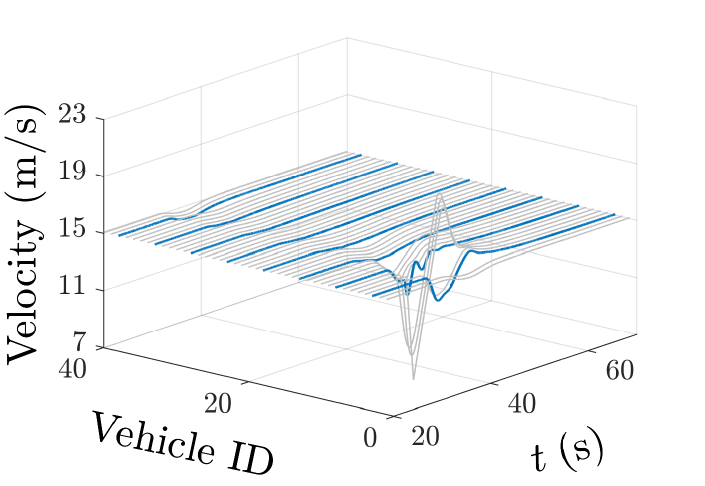}
		\includegraphics[scale=0.35]{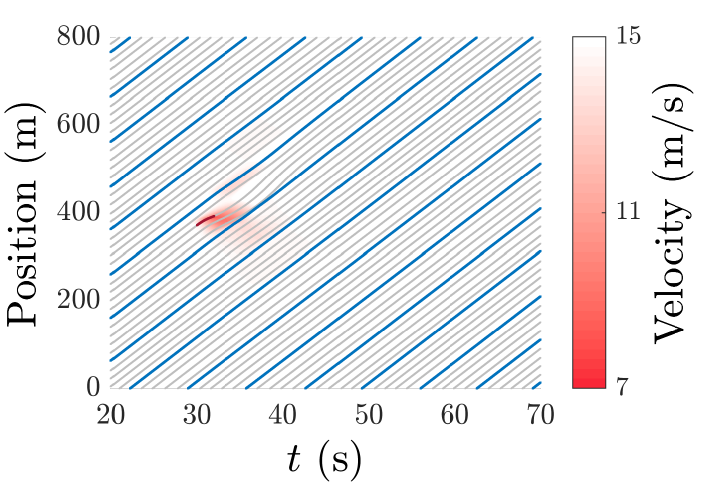}
	}
	\vspace{-2mm}
	\caption{Trajectory and velocity profiles of each vehicle when the perturbation happens at the 5th vehicle ($n=40,k=8$). In each panel, blue curves and gray curves represent the trajectories or velocity profiles of the AVs and the HDVs, respectively. The AVs are organized into a platoon in (a), while the AVs are distributed uniformly in (b).}
	\label{Fig:Simulation_Trajectory}
	\vspace{-2mm}
\end{figure}

The cooperative controller in Section~\ref{Sec:OptimalController} is adopted for the two formations, and the specific feedback gain $K_S$ is calculated based on the nominal parameter setup in OVM, with the parameter values in the performance output~\eqref{Eq:Output_Optimal} chosen as $\gamma_{s}=0.03, \gamma_{v}=0.15, \gamma_{u}=0.1$ (recall that upon this setup, the uniform distribution has an apparently larger probability to be the optimal formation, shown in Fig.~\ref{Fig:FormationResult}(c)(d)).  Furthermore, we consider two other practical factors~\cite{wang2018infrastructure}: 
a communication delay of $0.2  \, \mathrm{s} $ is incorporated~\cite{xing2019compensation,zhang2020semi}; a third-order model is employed to depict the longitudinal dynamics of the AVs~\cite{zheng2015stability}
\begin{equation} \label{Eq:FirstOrderDynamics}
    \tau_i \dot{a}_i(t) + a_i(t) = u_i(t).
\end{equation}
This assumes a first-order inertia process from the control input (\ie, desired acceleration) $u_i(t)$ to the real acceleration $a_i(t)$. The inertial delay $\tau_i$ is set to $\tau_i = 0.2+\mathbb{U}[-0.1,0.1]$. 
Finally, to guarantee driving safety and avoid crashes, we assume that all vehicles are equipped with a standard automatic emergency braking strategy, described as follows
\begin{equation}
	\dot{v}_{i}(t)=a_{\min },\; \mathrm { if }\, \frac{v_{i}^{2}(t)-v_{i-1}^{2}(t)}{2 s_{i}(t)} \geq \left|a_{\min }\right|,
\end{equation}
where the maximum acceleration and deceleration rates are set to $a_{\max }=2\, \mathrm{m} / \mathrm{s}^{2}, a_{\min }=-5\, \mathrm{m} / \mathrm{s}^{2}$, respectively. All vehicles have a maximum velocity limit of $30\,\mathrm{m/s}$, similar to the OVM.

Here, we consider a scenario where one single vehicle suffers from a sudden and rapid perturbation, which often occurs at lane changes or merging lanes. Specifically, we assume that the traffic flow starts with an equilibrium state, and at $t=30\,\mathrm{ s}$, one vehicle begins to brake at $-5\,\mathrm{ m}/\mathrm{s}^2$ for $1.5\, \mathrm{s}$. Fig.~\ref{Fig:Simulation_Trajectory} shows the vehicle trajectories when the perturbation happens at the 5th vehicle. As can be clearly observed in Fig.~\ref{Fig:Simulation_Trajectory_Platoon}, when the AVs are organized into a platoon, the traffic wave persists until it reaches the position of the platoon. Hence, when the perturbation happens ahead of the platoon and close to the platoon leader, the platoon achieves an impressive performance: it quickly dissipates the perturbation, and stops it from continuing to propagate upstream. However, when the perturbation is introduced somewhere else, the platoon fails to dampen the traffic waves in a short time and the uniform distribution behaves much better under these conditions.

\begin{figure}[t]
	\centering
	\subfigure[]
	{ \includegraphics[scale=0.4]{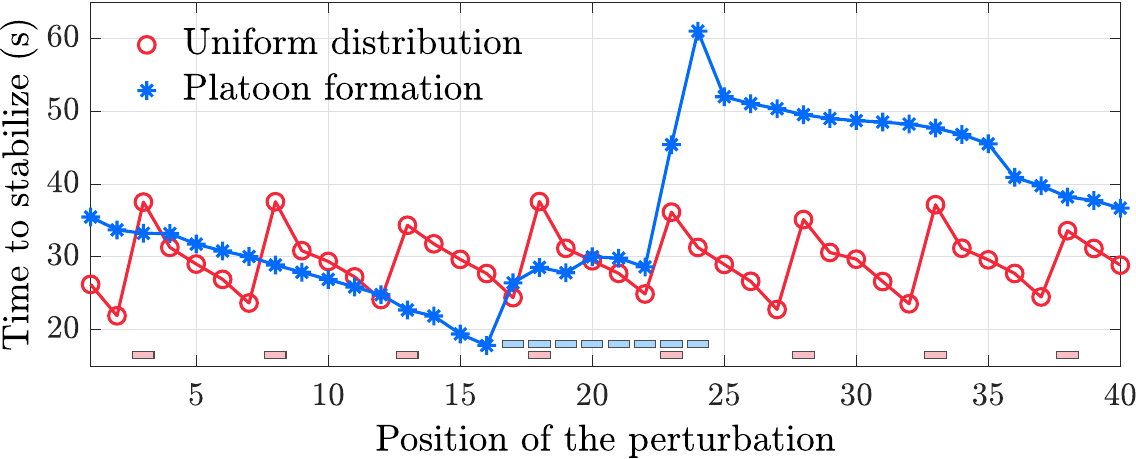}
	}\\
	\vspace{-2mm}
	\subfigure[]
	{
		\includegraphics[scale=0.4]{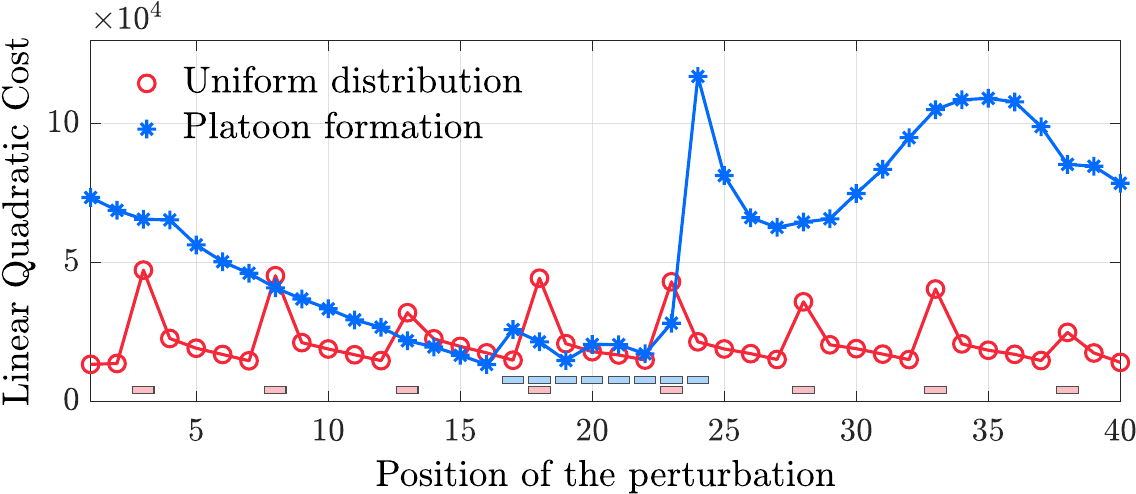}
	}
	\vspace{-2mm}
	\caption{Performance comparison at different positions of the perturbation ($n=40$, $k=8$). The small blue rectangles and the small red rectangles represent the location of AVs under a platoon formation ($S=\{17,18,19,21,21,22,23,24\}$) or a uniform distribution ($S=\{3,8,13,18,23,28,33,38\}$), respectively. (a) The stabilizing time of the traffic flow. (b) The linear quadratic cost, defined as $\int_{t=0}^{\infty} \left(x^{\tr}(t) Q x(t)+u^{\tr}(t) R u(t)\right) dt$ with $Q$ and $R$ taking the same value as those in~\eqref{Eq:Output_Optimal}.}
	\label{Fig:Simulation_Comparison}
	\vspace{-2mm}
\end{figure}

The comparison of two specific performance metrics at various positions of the perturbation in Fig.~\ref{Fig:Simulation_Comparison} validates this observation. It is evident to see that uniform distribution achieves a better performance than platoon formation in most cases, with only a few exceptions where the perturbation happens close to the platoon leader. This result indicates that platooning indeed has a great capability in dissipating traffic perturbations when the perturbation happens immediately ahead. Nevertheless, it is highly possible that perturbations happen somewhere else at a low penetration rate. In this case, other formations of AVs, \eg, uniform distribution, have a greater potential in reducing undesired instabilities and improving travel efficiency for the collective traffic flow.

\section{Discussions and Conclusions}
\label{Sec:Conclusion}

In this paper, we have introduced a set-function approach to describe the performance of mixed traffic systems with an explicit consideration of the cooperative formation of multiple AVs. The stability invariance property and diminishing improvement property of noncooperative formation under classical ACC strategies have been revealed. We have also formulated a set function optimization problem to investigate the optimal formation for AVs in mixed traffic flow. Considering the cooperative optimal control strategy and the resulting $\mathcal{H}_2$ performance, we reveal two predominant optimal formations for AVs: uniform distribution and platoon formation. Our results indicate that when HDVs have a poor string stability behavior, the prevailing vehicle platooning is not a suitable choice, which might even have the least potential in mitigating traffic perturbations. The results from nonlinear traffic simulation also support our findings.

\subsection{The Role of Optimal Formation}


Beyond the prevailing platoon formation, our set-function optimization approach and extensive numerical studies  have revealed huge potential for other possible formations of AVs in mixed traffic. Indeed, platoon formation and uniform distribution are two predominant patterns for the optimal formation; furthermore, the same formation might be the optimal choice in certain cases or the worst choice in others, depending on the HDVs' car-following behavior and traffic equilibrium states. Practical traffic flow typically has time-varying equilibrium states; individual HDVs' dynamics may be time-varying in different situations, and lane-changing behaviors also happen. Then, one natural question is that \emph{whether it is always necessary to pursue the globally ``optimal formation'' of AVs}. 

This question definitely deserves further investigation, especially considering required formation changing maneuvers, such as leaving, merging, and splitting. Our intuition is that active formation changes are unnecessary. 
Frequent formation changing maneuvers cause additional traffic disturbances, negatively impacting traffic flow~\cite{mena2018impact,deng2016general}. 
In this paper, our results confirm the benefits and potential of AVs with cooperative formation and control in mixed traffic, beyond platooning. Instead of performing formation changing maneuvers to achieve the platooning or another specific pattern, it might be more feasible to maintain their natural formation in mixed traffic. This natural formation can naturally change considering that the surrounding HDVs could change lanes. Then, it is very beneficial to apply cooperative control strategies of AVs according to the current formation (\eg, the redesigned optimal controller in Section~\ref{Sec:Re-designed}). 

\subsection{Limitations and Future Directions}

Several limitations in our research deserves further investigation. For noncooperative formation, we considered an independently designed ACC controller as a particular example, and then provided some theoretical analysis for its role in mixed traffic, \eg, stability and submodularity. It will be interesting to consider re-designed ACC controllers in different formations. 
For cooperative formation, we assumed a centralized control framework that requires access to global traffic states. Hence, V2V/V2I is required for all involved vehicles. Technologies, such as edge/cloud computing, are essential to implement and maintain various formations our current cooperative control. How to incorporate these technologies efficiently deserves further investigations. It is of great interest to utilize limited traffic information for designing distributed control from either a microscopic perspective~\cite{wang2020controllability,zheng2015stability} or a macroscopic perspective~\cite{abadi2014traffic,treiber2013traffic}.
    
Finally, our research is based on a closed single-lane ring road traffic system~\cite{cui2017stabilizing,zheng2020smoothing, giammarino2020traffic,stern2018dissipation}. By modifying the system model~\eqref{Eq:SystemModel}, our theoretical formulation for cooperative formation can be extended to an open straight road scenario, but the analysis on the formation of AVs is left for future work. It is also interesting to carry out large-scale traffic simulations to investigate the potential of different formations in more practical mixed traffic flow, \textit{e.g.}, utilizing high-fidelity traffic simulation packages such as SUMO~\cite{lopez2018microscopic} and VISSIM~\cite{fellendorf2010microscopic}, where more complex traffic scenarios and common traffic behaviors (\eg, lane-changing) could be incorporated.

%



\section*{Acknowledgment}
The authors would like to thank the anonymous reviewers and Associate Editor for their feedback, which helped improve the quality of our work. 

\appendices

\section*{Appendices}
\addcontentsline{toc}{section}{Appendices}
\renewcommand{\thesubsection}{\Alph{subsection}}


In this appendix, we provide some auxiliary results to the main text, including interpretations of $\mathcal{H}_2$ performance, the proof of stability invariance property, and a numerical algorithm for examining the submodularity.   

\subsection{Interpretations of $\mathcal{H}_2$ Performance}
\label{App:H2}
Given a general stable system $\dot{x}=A x+H w$ with output $z=C x$, the $\mathcal{H}_{2}$ norm of its transfer function $\mathbf{G}$ from $w$ to $z$ has intuitive physical interpretations~\cite{skogestad2007multivariable}:

\emph{1) Energy of the impulse response}: Denote $z_{i}$ as the performance output when the $i$-th input channel of the system is fed an impulse and $N$ as the total number of the input channels. The $\mathcal{H}_{2}$ performance quantifies the sum of the energy of the impulse response
	\begin{equation}
	\Vert \mathbf{G} \Vert _{2}^{2}=\sum_{i=1}^{N} \int_{0}^{\infty} z_{i}^{\tr}(t) z_{i}(t) d t.
	\end{equation}
	For the traffic system~\eqref{Eq:SystemModel}, common traffic bottlenecks, such as lane changing, merging and on-ramps, could lead to stop-and-go traffic waves. To model such scenarios, one can assume that the vehicle accelerations are subject to an impulse disturbance. In this case, $\mathcal{H}_{2}$ performance quantifies the sum of the velocity and spacing deviations caused by such disturbance.
	
 \emph{2) Expected power of the response to white noise}: When the disturbance signal $\omega$ is a white second-order process with unit covariance, the $\mathcal{H}_{2}$ performance measures the
	expected steady-state output
	\begin{equation}
	\Vert \mathbf{G} \Vert_{2}^{2}=\lim _{t \rightarrow \infty} \mathbb{E}\left(z^{\tr}(t) z(t)\right).
	\end{equation}
	Besides traffic bottlenecks, it is known that traffic waves could also emerge from the collective dynamics of drivers' uncertain behaviors. To depict this scenario, one can force a persistent stochastic noise at each vehicle's acceleration, and the $\mathcal{H}_{2}$ performance is the expectation of the steady variation of velocity and spacing deviations.

\subsection{Stability Invariance}
 \label{App:ProofStability}
 
 The notion of stability in Section~\ref{Sec:StabilityInvariance} refers to the Lyapunov sense. The formation definition is as follows.
 
 \begin{definition} [Lyapunov stability \cite{skogestad2007multivariable}]
 	Consider a dynamical system $\dot{x}=f(x(t))$. The equilibrium point $x_{e}$ is said to be Lyapunov stable, if $\forall \epsilon>0$, there exists $\delta>0$ such that, if $\Vert x(0)-x_{e}\Vert_2<\delta$, then $\Vert x(t)-x_{e} \Vert_2<\delta$ for every $t \geq 0$. The equilibrium point $x_{e}$ is said to be asymptotically stable, if it is Lyapunov stable, and there exists $\delta>0$ such that, if $\Vert x(0)-x_{e}\Vert_2<\delta$, then $\lim _{t \rightarrow \infty}\Vert x(t)-x_{e}\Vert_2=0$.
 \end{definition}

In the following, we present the proof of the stability invariance property revealed in Section~\ref{Sec:StabilityInvariance}. 

\begin{IEEEproof}[Proof of Theorem~\ref{Th:Stability}]
	The distribution of the eigenvalues of $\widehat{A}_{S}$ is characterized by
	$$\mathrm{det}\left(\lambda I_{2 n}-\widehat{A}_{S}\right)=0,$$
	where $\mathrm{det}(\cdot)$ denotes the determinant of a matrix and
	$$
	\lambda I_{2 n}-\widehat{A}_{S}=\begin{bmatrix}\lambda I_{n} & -M_{1} \\ -\alpha_{1} I_{n}+k_{s} D_{S} & \lambda I_{n}-M_{2}+k_{v} D_{S}\end{bmatrix}.
	$$
	Given a block matrix
	$$
	M=\begin{bmatrix}A & B \\ C & D\end{bmatrix},
	$$
	we have that~\cite[Theorem 3]{silvester2000determinants}
	$$
	\mathrm{det} M=\mathrm{det}(A D-C B),\; \mathrm{if}\, A C=C A.
	$$
	Thus, we have
	\begin{equation} \label{Eq:EigenvalueDerivation}
	\begin{aligned}
	&\mathrm{det}\left(\lambda I_{2 n}-\widehat{A}_{S}\right)\\
	=&\mathrm{det}\left(\lambda^{2} I_{n}-\lambda M_{2}+\lambda k_{v} D_{S}-\alpha_{1} M_{1}+k_{s} D_{S} M_{1}\right)\\
	\end{aligned}
	\end{equation}
	
	\begin{equation*}
	\begin{aligned}
	=&\mathrm{det}\left(\begin{bmatrix}g_{1} & &\cdots & -h_{1} \\ -h_{2} & g_{2} \\ &\ddots & \ddots& \\ & & -h_{n} & g_{n}\end{bmatrix}\right)\\
	=&\prod_{i=1}^{n} g_{i}-\prod_{i=1}^{n} h_{i},
	\end{aligned}
	\end{equation*}
	with
	$$
	\begin{aligned}
	g_{i}&=\lambda^{2}+\left(\alpha_{2}+k_{v} \delta_{i}\right) \lambda+\alpha_{1}-k_{s} \delta_{i},\\
	h_{i}&=\lambda \alpha_{3}+\alpha_{1}-k_{s} \delta_{i}, \quad i=1,\ldots,n.
	\end{aligned}
	$$
	Considering $|S|=k$ and substituting the definition of $\delta_{i}$ in~\eqref{Eq:Delta}
	to~\eqref{Eq:EigenvalueDerivation}, we have
	\begin{equation} \label{Eq:EigenvalueEquation}
	\begin{aligned}
	&\left(\lambda^{2}+\alpha_{2} \lambda+\alpha_{1}\right)^{n-k}\left(\lambda^{2}+\left(\alpha_{2}+k_{v}\right) \lambda+\alpha_{1}-k_{s}\right)^{k}\\
	&-\left(\lambda \alpha_{3}+\alpha_{1}\right)^{n-k}\left(\lambda \alpha_{3}+\alpha_{1}-k_{s}\right)^{k}=0.
	\end{aligned}
	\end{equation}
	It is clear that equation~\eqref{Eq:EigenvalueEquation} relies only on the number of AVs $k$, \ie, $|S|$ (the penetration rate), and is independent to the explicit elements of $S$. Thus, the distribution of eigenvalues of $\widehat{A}_S$ remains unchanged when $|S|$ is fixed.
\end{IEEEproof}

\begin{algorithm}[b]
	\caption{Examine the submodularity of $J(S)$}
	\label{Alg:Submodularity}
	\begin{algorithmic}[1]
		\Require
		$J(S)$, $\Omega$, $n$, number of experiments $N$;
		\Ensure
		result of submodularity $submodular$;
		\State Initialize $k \leftarrow 0$, $submodular\leftarrow true$;
		\While{$k<N$ and $submodular$}
		\State  Choose $a \in \Omega$ $(a \neq 1)$ randomly;
		\State $S_{1} \leftarrow\{a\}$, $i \leftarrow 1$;
		\State	$\Delta_{J}\left(1 | S_{1}\right) \leftarrow J\left(S_{1} \cup\{a\}\right)-J\left(S_{1}\right)$;
		\While {$i<n-1$}
		\State Choose $a \in \Omega \backslash S_{i}$ $(a \neq 1)$ randomly;
		\State	$S_{i+1} \leftarrow S_{i} \cup\{a\}$, $i \leftarrow i+1$;
		\State 	$\Delta_{J}\left(1 | S_{i}\right) \leftarrow J\left(S_{i} \cup\{a\}\right)-J\left(S_{i}\right)$;
		\EndWhile
		\If {$\left\{\Delta_{J}\left(1 | S_{i}\right)\right\}$ is not non-increasing}
		\State $submodular \leftarrow  false$;
		\EndIf
		\State $k \leftarrow k+1$;
		\EndWhile
		\State return $submodular$.
	\end{algorithmic}
\end{algorithm}
\vspace{-3mm}

\subsection{Examination of Submodularity}
\label{App:Submodularity}

To examine the submodularity of the performance value function $J(S)$, we design an algorithm based on Monte-Carlo simulations; see Algorithm~\ref{Alg:Submodularity}. The main idea is based on Lemma~\ref{Lem:SubmodularityValidation}. Given a specific parameter setup, first we generate an extensive number of random sequences of the marginal improvement $\left\{\Delta_{J}\left(1 | S_{i}\right)\right\}$, $i=1,2, \ldots, n,$ which satisfy~\eqref{Eq:RandomSequence}. If all random sequences $\left\{\Delta_{J}\left(1 | S_{i}\right)\right\}$ are non-increasing, it can be conjectured that $J(S)$ is submodular under the parameter setup. With one one counterexample identified, by contrast, it can be concluded immediately that $J(S)$ is not submodular.

\ifCLASSOPTIONcaptionsoff
  \newpage
\fi



%

\balance

\bibliographystyle{IEEEtran}
\bibliography{IEEEabrv,mybibfile}

%

\begin{IEEEbiography}[{\includegraphics[width=1in,height=1.25in,clip,keepaspectratio]{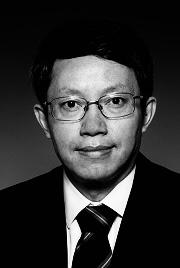}}]{Keqiang Li} received the B.Tech. degree from Tsinghua University of China, Beijing, China, in 1985, and the M.S. and Ph.D. degrees in mechanical engineering from the Chongqing University of China, Chongqing, China, in 1988 and 1995, respectively.
	
He is currently a Professor with the School of Vehicle and Mobility, Tsinghua University. His main research areas include automotive control system, driver assistance system, and networked dynamics and control. He is leading the national key project on ICVs (Intelligent and Connected Vehicles) in China. Dr. Li has authored more than 200 papers and is a co-inventor of over 80 patents in China and Japan.
	
Dr. Li has served as Fellow Member of Society of Automotive Engineers of China, editorial boards of the \emph{International Journal of Vehicle Autonomous Systems}, Chairperson of Expert Committee of the China Industrial Technology Innovation Strategic Alliance for ICVs (CAICV), and CTO of China ICV Research Institute Company Ltd. (CICV). He has been a recipient of Changjiang Scholar Program Professor, National Award for Technological Invention in China, etc.
\end{IEEEbiography}

\begin{IEEEbiography}[{\includegraphics[width=1in,height=1.25in,clip,keepaspectratio]{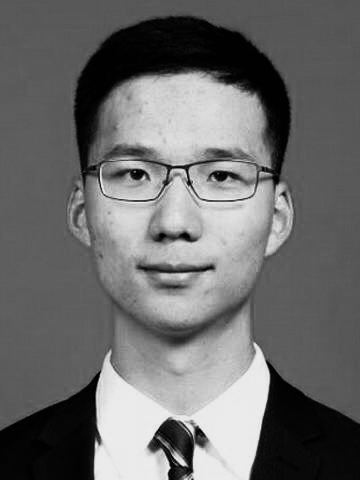}}]{Jiawei Wang} (Graduate Student Member, IEEE) received the B.E. degree from Tsinghua University, Beijing, China, in 2018. He is currently a Ph.D. student in mechanical engineering with the School of Vehicle and Mobility, Tsinghua University. He is currently also a visiting Ph.D. student with the Automatic Control Laboratory at the École Polytechnique Fédérale de Lausanne (EPFL). His research interests include connected automated vehicles, distributed control and optimization, and data-driven control. He was a recipient of the National Scholarship in Tsinghua University. He received the Best Paper Award at the 18th COTA International Conference of Transportation Professionals. 
\end{IEEEbiography}
	
\begin{IEEEbiography}[{\includegraphics[width=1in,height=1.25in,clip,keepaspectratio]{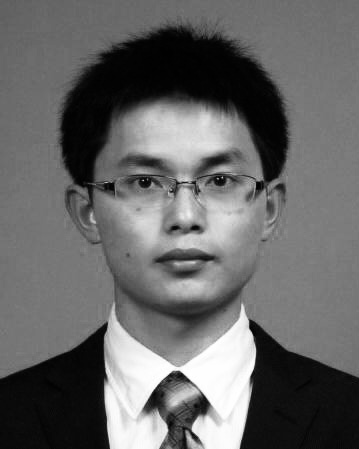}}]{Yang Zheng}(Member, IEEE) received the B.E. and M.S. degrees from Tsinghua University, Beijing, China, in 2013 and 2015, respectively, and the D.Phil. (Ph.D.) degree in Engineering Science from the University of Oxford, U.K., in 2019.
	
He is currently an assistant professor with the Department of Electrical and Computer Engineering, UC San Diego. He was a research associate at Imperial College London and was a postdoctoral scholar in SEAS and CGBC at Harvard University. His research interests focus on learning, optimization, and control of network systems, and their applications to autonomous vehicles and traffic systems.

Dr. Zheng was a finalist (co-author) for the Best Student Paper Award at the 2019 ECC. He received the Best Student Paper Award at the 17th IEEE ITSC in 2014, and the Best Paper Award at the 14th Intelligent Transportation Systems Asia-Pacific Forum in 2015. He was a recipient of the National Scholarship, Outstanding Graduate in Tsinghua University, the Clarendon Scholarship at the University of Oxford, and the Chinese Government Award for Outstanding Self-financed Students Abroad. Dr. Zheng won the 2019 European PhD Award on Control for Complex and Heterogeneous Systems.
\end{IEEEbiography}



\end{document}